\documentclass[preprint,3p]{elsarticle}
\usepackage[fleqn]{amsmath}
\usepackage{amssymb,amsthm}
\usepackage{mathtools}
\usepackage{mathrsfs}
\usepackage{bm}
\usepackage{slashed}	%feynman slashed symbols
\usepackage{cancel}
\usepackage{soul}
\usepackage{color}
\usepackage{xcolor}
\usepackage[pdftex,hypertexnames=false,linktocpage=true]{hyperref}
\hypersetup{colorlinks=true,linkcolor=red,anchorcolor=darkgreen,citecolor=green!50!blue,filecolor=darkgreen,urlcolor=green,
bookmarksnumbered=true,
pdfview=FitB
}
\usepackage{lineno}
\biboptions{sort&compress}
\usepackage{verbatim}
\usepackage{fullpage}
\usepackage{graphicx}
\usepackage{tikz}
\usepackage{subfig} 	% a set of figures
\usepackage{relsize}	%rescalable math
\usepackage{float}

\usepackage{array}
\usepackage{booktabs}
\usepackage{arydshln}
\usepackage{lscape}
\usepackage{multirow}
\usepackage{pdfpages}

\def\dd{{\mathrm{d}}}

\mathchardef\-="2D

\newcommand{\half}[1][1] {\mathsmaller{\frac{#1}{2}}}

\makeatletter

\newcommand{\Rmnum}[1]{\expandafter\@slowromancap\romannumeral #1@}
\makeatother
\colorlet{darkgreen}{green!60!black}
\colorlet{brightyellow}{yellow!75!red}
\colorlet{orange}{red!50!yellow}
\colorlet{darkblue}{blue!60!black}
\colorlet{darkred}{red!80!black}
\colorlet{greenblue}{green!50!blue}

\begin{document}
\begin{frontmatter}

%title and authors
\title{Heavy Quarkonium in a Holographic Basis}

\author[isu]{Yang~Li}\corref{c1}
\ead{leeyoung@iastate.edu}

\author[isu]{Pieter~Maris}
\ead{pmaris@iastate.edu}

\author[imp]{Xingbo~Zhao}
\ead{xbzhao@impcas.ac.cn}

\author[isu]{James~P.~Vary}
\ead{jvary@iastate.edu}

\cortext[c1]{Corresponding author}

\address[isu]{Department of Physics and Astronomy, Iowa State University, Ames, IA 50011, USA}
\address[imp]{Institute of Modern Physics, Chinese Academy of Sciences, Lanzhou 730000, China}

%abstract
\begin{abstract}

We study the heavy quarkonium within the basis light-front quantization approach.
We implement the 
one-gluon exchange interaction and a confining potential inspired by light-front holography. We adopt the holographic 
light-front wavefunction (LFWF) as our basis function and solve the non-perturbative dynamics by diagonalizing the 
Hamiltonian matrix. We obtain the mass spectrum for charmonium and bottomonium. With the obtained LFWFs, 
we also compute the decay constants and the charge form factors for selected eigenstates. The results are compared with the 
experimental measurements and with other established methods. 

\end{abstract}

\end{frontmatter}

\section{Introduction} \label{sect.1}

Describing hadrons from quantum chromodynamics (QCD) remains a fundamental challenge in nuclear physics. Inspired by the discovery of a
remarkable gauge/string duality \cite{Maldacena98.231}, holographic QCD models, most notably the AdS/QCD \cite{Karch06.015005}, have been
proposed as analytic semi-classical approximations to QCD (for a recent review, see Ref.~\cite{Kim13.55}). In light of these
phenomenological successes, as well as the recent progress in the \textit{ab initio} nuclear structure calculations
\cite{Barrett13.131,Caprio15.1541002,Maris10.97,Aktulga14.2631}, the basis light-front quantization (BLFQ) \cite{Vary10.035205} has been
developed as a non-perturbative approach to address QCD bound-state problems from first principles. 

BLFQ is based on the Hamiltonian formalism in light-front dynamics (LFD, \cite{Bakker13.165}) in Minkowski space. The central task of the
Hamiltonian approach is to diagonalize the QCD Hamiltonian operator, 
  \begin{linenomath*}
  \begin{equation}\label{eqn:LFHamiltonian}
   P^+ \hat P^- |\psi_h\rangle = M^2_h  |\psi_h\rangle.
  \end{equation}
  \end{linenomath*}
Here $P^\pm = P^0 \pm P^3$ is the longitudinal momentum and the light-front quantized Hamiltonian operator, respectively. 
The eigenvalues directly produce the invariant-mass spectrum. The eigenfunctions, known as the light-front wavefunctions (LFWFs), play a
pivotal role in the study of the hadron structures in deep inelastic scattering (DIS)
\cite{Lepage80.2157} and deeply virtually Compton
scattering (DVCS) \cite{Brodsky01.99}. 
In the Fock space expansion, Eq.~(\ref{eqn:LFHamiltonian})
becomes a relativistic quantum many-body problem and can be solved by constructing and diagonalizing the many-body Hamiltonian matrix (see,
e.g., \cite{Brodsky98.299} for a review). 

The advantages of LFD are made explicit by BLFQ which can employ an arbitrary single-particle basis subject to completeness and
orthonormality. By adopting a single-particle AdS/QCD basis, BLFQ naturally extends the AdS/QCD LFWFs to the multi-particle Fock
sectors
\cite{Vary10.035205}. Furthermore, this basis preserves all the kinematical symmetries of the 
full Hamiltonian \cite{Li13.136,Maris13.321}. Such choice is in parallel with the no-core shell model (NCSM) used in non-relativistic
quantum many-body theory \cite{Barrett13.131}. State-of-the-art computational tools developed in the many-body theory can be used to
address the QCD eigenvalue problem \cite{Vary09.012083}. BLFQ has been applied successfully to a range of
non-perturbative problems, including the electron anomalous magnetic moment \cite{Heli10.061603,Zhao14.65}, non-linear Compton scattering
\cite{Zhao13.065014,Zhao:2013jia} and the positronium spectrum
\cite{Wiecki15.105009,Vary15}.
In this paper, we apply the BLFQ approach to the heavy quarkonium.

Working with the full QCD Hamiltonian is a formidable task. In practice, we truncate the Fock space to a finite number 
of particles. The leading-order truncation $|q\bar{q}\rangle + |q\bar{q}g\rangle$ introduces the one-gluon exchange which produces 
correct short-distance physics as well as the spin-dependent interaction needed for the fine and hyperfine structures.
The Abelian version of this interaction was extensively used in the literature
\cite{Kaluza91.2968,Krautgartner92.3755,Trittmann:1997xz,Lamm14.125003,Wiecki15.105009} to calculate the QED bound-state spectrum in LFD.
However, the one-gluon exchange itself is not sufficient to reproduce the hadron spectrum since confinement is also needed. Holographic QCD
provides an appealing approximation to confinement.

Heavy quarkonium is an ideal laboratory for studying non-perturbative aspects of QCD and their interplay with the perturbative physics
\cite{Brambilla11.1534}. Conventional theoretical tools include the non-relativistic potential models (NRPMs)
\cite{Appelquist78.387,Godfrey85.189},
non-relativistic QCD (NRQCD) \cite{Brambilla05.1423}, heavy quark effective field theory \cite{Neubert94.259}, 
Dyson-Schwinger Equations (DSE) \cite{Bhagwat08.025203,Blank11.096014,Hilger15.034013,Fischer15.10}, and Lattice QCD
\cite{McNeile04.100}. 
The recent discoveries of tetraquark \cite{tetraquark} and pentaquark \cite{pentaquark} states have renewed interests in the
theoretical investigation of heavy quarkonium. 
Extensive data on heavy quarkonium have been produced by experimental facilities, such as Belle, CLEO and LHC. 
 
Numerous light-front phenomenologies have been developed for heavy quarkonium (see e.g. 
\cite{Singh86.47,Brisudova96.1831,Brisudova97.1227,Glazek06.105015,Choi07.073016,Branz10.074022,Gutsche14.096007} and the
references therein). Our approach shares some similarity with these models. Yet, there are also major differences. First of all,
our approach employs holographic QCD (confining interaction) and realistic LFQCD (one-gluon exchange).
Secondly and most importantly, we solve quarkonium as a two-body bound-state problem using a Hamiltonian method that is applicable to
arbitrary many-body bound states, once the (effective) Hamiltonian and the basis space are specified. We exploit the fact that
BLFQ is developed as a flexible computational platform for relativistic strong interaction many-body bound-state problems
\cite{Vary10.035205,Vary09.012083}, designed to deal with general Hamiltonians, 
realistic or phenomenological.

Our goal in this work can be simply stated: we aim to improve the light-front holographic QCD results
\cite{Brodsky15.1} by including a realistic one-gluon exchange interaction.  Computationally, we intend
to lay the foundation for the extension to higher Fock sectors.

 \section{Effective Hamiltonian} \label{sect.2}
 \subsection{Phenomenological confinement} \label{sect.2.1}

Our effective Hamiltonian consists of the holographic QCD Hamiltonian and the one-gluon exchange. We adopt the light-front
AdS/QCD soft-wall (SW) Hamiltonian for the first part \cite{Brodsky09.081601}. This simple model gives a reasonable description of the
hadron
spectrum and structures (see Ref.~\cite{Brodsky15.1} for a review).
 Its effective ``light-cone'' Hamiltonian reads, 
 \begin{linenomath*}
\begin{equation}\label{eqn:Heff}
 H_\textsc{sw} \equiv P^+\hat P^-_\textsc{sw} - \bm P^2_\perp 
 = \frac{\bm k^2_\perp}{x(1-x)} + \kappa^4 x(1-x)\bm r^2_\perp,
\end{equation}
\end{linenomath*}
where, $x = p^+_q/P^+$ is the longitudinal momentum fraction of the quark, $\bm k_\perp = \bm p_{q\perp} - x \bm P_\perp$ is the 
relative transverse momentum, and $\bm r_\perp$ is the
transverse separation of the partons. 
$\kappa$ is the strength of the confining potential. Note that the ``light-cone
Hamiltonian'' has mass squared dimension, whose eigenvalues are the squared invariant masses.
Following Brodsky and de T\'eramond \cite{Brodsky09.081601}, it is convenient to introduce the holographic coordinate $\bm \zeta_\perp =
\sqrt{x(1-x)}\bm r_\perp$, and its conjugate $\bm q_\perp = {\bm k_\perp}/{\sqrt{x(1-x)}} \equiv -i\nabla_{\zeta_\perp}$.
In light-front holography, $\zeta_\perp$ is mapped to the fifth coordinate $z$ of the AdS space. 
In these coordinates, $H_\textsc{sw}$ is a harmonic oscillator (HO),
\begin{linenomath*}
\begin{equation}\label{eqn:soft_wall}
 H_\textsc{sw} = \bm q^2_\perp + \kappa^4 \bm \zeta^2_\perp.
\end{equation}
\end{linenomath*}
Its eigenvalues follow the Regge trajectory $M^2 = 2\kappa^2 (2n + |m| + 1)$. Its eigenfunctions are 2D HO functions in the holographic
variables, 
\begin{linenomath*}
\begin{equation}
\phi_{nm}(\bm q_\perp) = e^{im\theta} \left(\frac{q_\perp}{\kappa} \right)^{|m|}
e^{-q^2_\perp/(2\kappa^2)} L_n^{|m|}(q^2_\perp/\kappa^2).
\end{equation}
\end{linenomath*}
Here $q_\perp = |\bm q_\perp|$, $\theta = \arg\bm q_\perp$, and $L_n^m(z)$ is the associated Laguerre polynomial. We
adopt
these functions as our basis. This basis has the advantage that in the many-body sector, it allows the exact factorization of the
center-of-mass motion in the single-particle coordinates. This is a very valuable property,
because the boson/fermion symmetrization/anti-symmetrization in the relative coordinates quickly becomes intractable, as the number of
identical particles increases \cite{Li13.136,Maris13.321}. For this work, however, we do not have identical particles in the $q\bar{q}$
sector and we will use the relative coordinate. In future extensions, as sea quarks and gluons are added, it may be more advantageous to
adopt single-particle coordinates.

The soft-wall Hamiltonian Eq.~(\ref{eqn:soft_wall}) is designed for massless quarks, and it is inherently 2-dimensional. For the 
heavy quarkonium systems, it should be modified to incorporate the quark masses and the longitudinal dynamics,
\begin{linenomath*}
 \begin{equation}
  H_\textsc{sw} \to H_\textsc{con} = \bm q^2_\perp + \kappa^4 \bm \zeta^2_\perp +  \frac{m_q^2}{x} + \frac{m_{\bar{q}}^2}{1-x} +
V_L(x).
 \end{equation}
\end{linenomath*}
Here $V_L$ is a longitudinal confining potential. Several longitudinal confining
potentials have been proposed 
\cite{Glazek11.1933,Trawinski14.074017,Chabysheva13.143}. Here we propose a new longitudinal confinement which shares 
features with others proposed,
\begin{linenomath*}
\begin{equation}\label{eqn:longitudinal_confinement}
V_{L}(x) = -\frac{\kappa^4}{(m_q+m_{\bar{q}})^2} \partial_x \big( x (1-x) \partial_x\big),
\end{equation}
\end{linenomath*}
where $\partial_x \equiv \left({\partial }/{\partial x}\right)_{\bm \zeta_\perp}$. This term combined with the mass term from the
kinetic energy forms a Sturm-Liouville problem,
\begin{linenomath*}
\begin{equation} \label{eqn:longitudinal_dynamics}
 - \partial_x \big( x(1-x) \partial_x \chi_l(x) \big) + \frac{1}{4}\Big( \frac{\alpha^2}{1-x} + \frac{\beta^2}{x} \Big) 
\chi_l(x) = \lambda_l^{(\alpha,\beta)} \chi_l(x), 
\end{equation}
\end{linenomath*}
where $\alpha = {2m_{\bar{q}}(m_q+m_{\bar{q}})}/{\kappa^2}$, $\beta = {2m_{q}(m_q+m_{\bar{q}})}/{\kappa^2}$.
The solutions of Eq.~(\ref{eqn:longitudinal_dynamics}) are analytically known in terms of the Jacobi polynomial $P^{(a,b)}_l(z)$, 
\begin{linenomath*}
\begin{equation}
 \chi_l(x) = x^{\half\alpha} (1-x)^{\half\beta} P^{(\alpha,\beta)}_l(2x-1).
\end{equation}
\end{linenomath*}
and form a complete orthogonal basis on the interval $[0,1]$. The corresponding eigenvalues are
\begin{linenomath*}
 \begin{equation}
  \lambda_l^{(\alpha,\beta)} = (l+\half(\alpha+\beta) )( l+\half(\alpha+\beta)+1 ). \quad (l=0, 1, 2,\cdots)
 \end{equation}
\end{linenomath*}

Comparing to other forms of longitudinal confinement, our proposal implements several attractive features.
First, the basis functions resemble the known asymptotic parton distribution $\sim x^\alpha (1-x)^\beta$ with $\alpha,\beta>0$
\cite{Antonuccio97.104}. 
This is our primary motivation for adopting the longitudinal confinement Eq.~(\ref{eqn:longitudinal_confinement}).
Second, the basis function is also analytically known, which brings numerical convenience. Third, in the non-relativistic limit
$\min\{m_q, m_{\bar q}\} \gg \kappa$, the longitudinal confinement sits on equal footing with the transverse
confinement, where together,
they form a 3D harmonic oscillator potential, 
\begin{linenomath*}
\begin{equation}
 V_\text{con} = \frac{m_qm_{\bar{q}}}{(m_q+m_{\bar{q}})^2} \kappa^4 \bm r^2,
\end{equation}
\end{linenomath*}
and rotational symmetry is manifest. This non-relativistic reduction also provides us an order-of-magnitude
estimate of the model parameters for our heavy quarkonium application. 
Fourth, in the massless limit $\max\{m_q, m_{\bar q}\} \ll \kappa$, the longitudinal mode stays in the ground
state and the
longitudinal wavefunction
$\chi_0(x) = \text{const}$. Thus we restore the massless model of Brodsky and de T\'eramond\footnote{
Note that in our normalization convention, the LFWFs differ from Brodsky et al.'s \cite{Brodsky15.1} by a factor $\sqrt{x(1-x)}$.}.

 \subsection{One-Gluon exchange} \label{sect.2.3}

Following Ref.~\cite{Wiecki15.105009} (cf. \cite{Kaluza91.2968, Krautgartner92.3755, Trittmann:1997xz,Lamm14.125003}), we
introduce the one-gluon exchange in LFD. In the momentum space, this term reads,
\begin{linenomath*}
\begin{equation}\label{eqn:one_gluon_exchange}
\langle \bm k'_\perp, x', s', \bar s' |  V_\text{g} | \bm k_\perp, x, s, \bar s \rangle = -\frac{4}{3} \times 
 \frac{4\pi \alpha_s}{Q^2} S_{s,\bar{s},s',\bar{s}'}(\bm k_\perp, x, \bm k'_\perp, x')
\end{equation}
\end{linenomath*}
where $S_{s,\bar{s},s',\bar{s}'}(\bm k_\perp, x, \bm k'_\perp, x') = \bar u_{s'}(\bm k'_\perp, x') \gamma_\mu u_{s}(\bm 
k_\perp, x) \bar v_{\bar{s}}(-\bm k_\perp, 1-x)\gamma^\mu 
v_{\bar{s}'}(-\bm k'_\perp, 1-x')$, and, 
$Q^2 
 = \frac{1}{2}\Big(\sqrt{\frac{x'}{x}}\bm k_\perp-\sqrt{\frac{x}{x'}}\bm k'_\perp\Big)^2 +
\frac{1}{2}\Big(\sqrt{\frac{1-x'}{1-x}}\bm k_\perp-\sqrt{\frac{1-x}{1-x'}}\bm k'_\perp \Big)^2 
+ \frac{1}{2}(x-x')^2\Big( \frac{m_q^2}{xx'}+\frac{m_{\bar q}^2}{(1-x)(1-x')}\Big) + \mu_g^2,$
is the average momentum transfer. 
We have also introduced a gluon mass $\mu_g$ to regularize the Coulomb singularity. This
singularity is integrable and 
does not carry physical significance. The gluon mass is used to improve the numerics and will be taken small compared to other
scales in this application.
In Eq.~(\ref{eqn:one_gluon_exchange}), we have taken the total momentum $\bm P_\perp = 0, P^+=1$ by virtue of the boost invariance
in LFD.

This one-gluon exchange introduces a logarithmic divergence 
\cite{Krautgartner92.3755, Mangin-Brinet03.055203,Lamm14.125003,Vary15}.
This can be seen from the one-gluon exchange kernel by transverse power counting. In particular, the leading power comes from 
the spin non-flip spinor matrix elements $S_{\uparrow\downarrow \uparrow\downarrow}$ and $S_{\downarrow \uparrow \downarrow\uparrow}$,
which contain terms proportional to $\bm k^2_\perp$ or $\bm k'^2_\perp$. Such terms do not vanish in the large momentum limit\footnote{
From the perturbative point of view, this should be canceled by a similar contribution in the cross 
ladder diagram, which is absent from our Fock space.}. In principle, this divergence can be handled by a proper renormalization
\cite{Perry91.4051,Glazek93.1599,Glazek93.5863,Hiller98.016006,Karmanov08.085028}. However, such a procedure is usually 
rather challenging or even impractical. In this work, we adopt a simple counterterm technique
\cite{Wiecki15.105009,Krautgartner92.3755,Trittmann:1997xz,Mangin-Brinet03.055203,Vary15} by replacing the spinor matrix elements
$S_{\uparrow\downarrow \uparrow\downarrow}$ and $S_{\downarrow \uparrow \downarrow\uparrow}$ with, 
\begin{linenomath*}
\begin{multline}
 S_{s,\bar{s},s',\bar{s}'}(\bm k_\perp, x, \bm k'_\perp, x') \to  S_{s,\bar{s},s',\bar{s}'}(\bm k_\perp, x, \bm k'_\perp, x') - 2\bm 
k^2_\perp \sqrt{\frac{x'(1-x')}{x(1-x)}} - 2\bm k'^2_\perp\sqrt{\frac{x(1-x)}{x'(1-x')}}. \\ (s, \bar s, s', \bar s' =
\;\uparrow\downarrow \uparrow\downarrow \text{or}
\downarrow
\uparrow \downarrow\uparrow)
\end{multline}
\end{linenomath*}
In its essence, this counterterm 
exactly removes the troubling $\bm k^2_\perp$ and $\bm k'^2_\perp$ terms in the spinor
matrix elements hence removing the UV divergence. 

\section{Basis light-front quantization} \label{sect.3}

In Sect.~\ref{sect.2}, we derived the effective light-cone Hamiltonian operator for quarkonium, which reads,
\begin{linenomath*}
\begin{equation}
 H_\text{eff} = \bm q^2_\perp + \kappa^4 \bm \zeta_\perp^2 + \frac{m_q^2}{x} + \frac{m_{\bar{q}}^2}{1-x} 
 - \frac{\kappa^4}{(m_q+m_{\bar{q}})^2}\partial_x \big(x(1-x) \partial_x \big) 
 + V_\text{g}.
\end{equation}
\end{linenomath*}
In this section, we will focus on solving the eigenvalue equation non-perturbatively, 
\begin{linenomath*}
\begin{equation}
H_\text{eff} \, | \psi^{J^{PC}}_{m_J} \rangle = M^2 | \psi^{J^{PC}}_{m_J} \rangle,
\end{equation}
\end{linenomath*}
to obtain the mass spectrum and the LFWFs $\psi^{J}_{m_J}(\bm k_\perp, x,s,\bar{s}) \equiv \langle\bm k_\perp, x,
s,\bar{s}|\psi^{J^{PC}}_{m_J}\rangle$. Here $J$, $P$, $C$ and $m_J$ are the total spin, parity, charge parity, and the magnetic
projection of the state, respectively.
Our strategy is to construct a finite-dimensional matrix from the effective Hamiltonian operator and then diagonalize it
numerically. To do this, we first adopt a basis expansion, 
\begin{linenomath*}
 \begin{equation}
  \psi^{J}_{m_J}(\bm k_\perp, x,s,\bar{s}) = \sum_{\mathclap{n,m,l}} \delta_{m_J,m+s+\bar s} \, \widetilde{\psi}^{J}_{m_J}(n,m,l,s,\bar{s})
\, \Phi_{nml}(\bm k_\perp/\sqrt{x(1-x)}, x).
\end{equation}
\end{linenomath*}
Here $\widetilde\psi^{J}_{m_J}(n,m,l,s,\bar{s})\equiv\langle n,m,l,s,\bar{s} | \psi^{J^{PC}}_{m_J} \rangle$ are the LFWFs in the BLFQ basis.
As mentioned above,
the basis functions $\Phi_{nml}(\bm q_\perp,x)\equiv\phi_{nm}(\bm q_\perp)\chi_l(x)$ are generated by the confining part of the Hamiltonian
--
the kinematic energy plus the confinement. As such,
the light-front holographic QCD results serve as our first approximation.
The matrix elements of the Hamiltonian within this basis are, 
\begin{linenomath*} 
\begin{multline} \label{eqn:Hamiltonian_matrix_element}
 \langle n',m',l',s',\bar{s}' | H_\text{eff} | n,m,l,s,\bar{s} \rangle = \langle n',m',l',s',\bar{s}' | V_\text{g} | n,m,l,s,\bar{s} 
\rangle \\
+ \Big[ (m_q + m_{\bar q})^2 + 2\kappa^2 (2n+|m|+l+\half[3]) + \frac{\kappa^4 }{(m_q+m_{\bar{q}})^2} l(l+1) \Big] 
\delta_{nn'}\delta_{mm'}\delta_{ll'} \delta_{ss'} \delta_{\bar{s}\bar{s}'}.
\end{multline}
\end{linenomath*}

We take advantage of the conservation of the angular momentum in the transverse plane, and choose a particular magnetic projection 
$m_J$:  
$ m + s + \bar s = m_J$.
In order to carry out practical calculations, we truncate the infinite basis space to a finite size by 
restricting the quantum numbers according to 
$ 2n + |m| + 1 \le N_{\max}, l \le L_{\max}$.
$N_{\max}$ controls the range of momenta covered by the harmonic oscillator basis as it is related to a
transverse IR regulator $\lambda_\textsc{ir} \sim b/\sqrt{N_{\max}}$ and a UV regulator $\Lambda_\textsc{uv} \sim
\sqrt{N_{\max}}b$ \cite{Coon:2012ab}, where $b$ is the HO basis scale and we have taken $b=\kappa$, unless otherwise 
stated. $L_{\max}$ controls the basis resolution in the longitudinal direction.

The relativistic bound states are identified by three quantum numbers $J^{PC}$. However, $P$ is broken by the finite basis truncation,
because
the parity transformation is dynamical in LFD. Instead, one can exploit the so-called
mirror parity $\hat P_x = \hat R_x(\pi) \hat P$ that is related to $P$ \cite{Soper72.1956},
\begin{linenomath*}
 \begin{equation}
\hat P_x |\psi^{J^{PC}}_{m_J}\rangle = (-1)^{J}{P}
|\psi^{J^{PC}}_{-m_J} \rangle. 
 \end{equation}
\end{linenomath*}
Then $P$ can be extracted from the LFWFs by, 
\begin{linenomath*}
\begin{equation}
(-1)^{J}P = \langle \psi^{J^{PC}}_{-m_J}|\hat R_x(\pi) \hat P | \psi^{J^{PC}}_{m_J}\rangle 
= \sum_{n,m,l,s,\bar{s}} (-1)^{m} \, \widetilde\psi^{J*}_{-m_J}(n,-m,l,-s, -\bar{s}) \widetilde\psi^J_{m_J}(n,m,l, s, \bar{s}).
\end{equation}
\end{linenomath*}
Similarly, $C$ can be obtained from the LFWFs by, 
\begin{linenomath*}
\begin{equation}
C = \langle \psi^{J^{PC}}_{m_J} | \hat C | \psi^{J^{PC}}_{m_J} \rangle = \sum_{n,m,l,s,\bar{s}} (-1)^{m+l+1} \, \widetilde\psi^{J*}_{m_J}(n,
m, l, s, \bar{s}) \widetilde\psi^J_{m_J}(n, m, l, \bar{s}, s).
\end{equation}
\end{linenomath*}
Rotation is also a dynamical symmetry in LFD. As a result, $J$ is not conserved and the mass degeneracy for different
magnetic projections $m_J$ is lifted by the Fock-space truncation. Nevertheless, in a non-relativistic system, such as the
heavy quarkonium, the discrepancy is small and
we can still extract $J$ by counting the multiplicity of the nearly-degenerate mass eigenstates. It is also instructive to assign states
the non-relativistic quantum numbers $n\,{}^{2S+1}\!L_J$, where $n$, $S$ and $L$ are the radial quantum number,
the spin and the orbital angular momentum, respectively. $L,S$ are related to parity and charge conjugation through 
$P=(-1)^{L+1}, C=(-1)^{L+S}$.
The radial quantum number $n$ can be deduced from the mass hierarchy of the spectrum. The total spin $S$,  though not an exact quantum
number, can be obtained from its expectation value, 
$\langle \psi^{J^{PC}}_{m_J} |\vec S^2|\psi^{J^{PC}}_{m_J}\rangle = S(S+1)$.
$L, S, J$ are constrained by the angular momentum addition 
$|L-S| \le J \le L+S$.
Reconstructing these quantum numbers allows us to identify the states and to compare with experimental data and with other methods. 

\section{Numerical Results} \label{sect.4}

In this work, we focus on charmonium and bottomonium, where the fermion masses are equal ($m_q = m_{\bar{q}}$) and heavy ($m_q \gtrsim
\kappa$). The Hamiltonian matrix element (Eq.~(\ref{eqn:Hamiltonian_matrix_element})) involves a four-dimensional integral (two in the
radial direction, two in the longitudinal direction). They are evaluated using Gauss quadratures. The number of the quadrature points
$N_\text{rad}$ ($N_\text{lfx}$) is taken to be at least twice $N_{\max}$ ($L_{\max}$).
Then the obtained matrix is diagonalized using LAPACK software \cite{lapack}. 
We fix the bottomonium coupling $\alpha_s(M_{b\bar b})=0.25$ and obtain the charmonium coupling $\alpha_s(M_{c\bar c})=0.3595$ using the
leading-order pQCD evolution of strong coupling ($N_f=5, M_{c\bar c}\simeq 3.5\,\mathrm{GeV}, M_{b\bar b}\simeq 9.5\,\mathrm{GeV}$). We
then fit the parameters $\kappa$, $m_q$ ($m_c$, $m_b$) by minimizing the root-mean-squared (r.m.s.) deviation from the experimental masses
\cite{pdg.2014} of the states below the $D \overline{D}$ or $B \overline{B}$ threshold. The details of the model parameters are summarized
in Table~\ref{tab:model_parameters}. In these calculations, we take the regulators $\mu_g=0.02\,\mathrm{GeV} \ll \kappa$,
$N_{\max}=L_{\max}=8, 16, 24$ ($N_{\max}=8$ is available in the supplemental materials), and the number of quadrature points $N_\text{rad} =
N_\text{lfx} = 64$.

Previous BLFQ study of positronium \cite{Wiecki15.105009, Vary15} shows that the continuum limit $N_{\max} \to \infty$, 
$L_{\max} \to
\infty$, $\mu_g \to 0$ can be reached through extensive calculation and successive extrapolations. However, we shall not investigate the 
$N_{\max}$, $L_{\max}$ and $\mu_g$ extrapolations in this paper because of the numerical efforts involved and also
the presence of several fitted parameters. Comparing the results with $N_{\max}=L_{\max}=16$ and $N_{\max}=L_{\max}=24$ in
Table~\ref{tab:model_parameters}, where the model parameters $\kappa, m_q$ were fitted separately and turned out to be very close
($\lesssim$ 1{\%} change). The r.m.s. deviations from the measured spectra are also comparable.
We also studied the trend of the mass eigenvalues with decreasing $\mu_g$ in the range of
$10^{-4} \,\mathrm{GeV} \le \mu_g \le 3\times 10^{-1} \,\mathrm{GeV}$ at fixed $N_{\max} = L_{\max}$, and found that the mass
eigenvalues are well converged with respect to $\mu_g$.

\begin{figure}
 \centering 
\includegraphics[width=.478\textwidth]{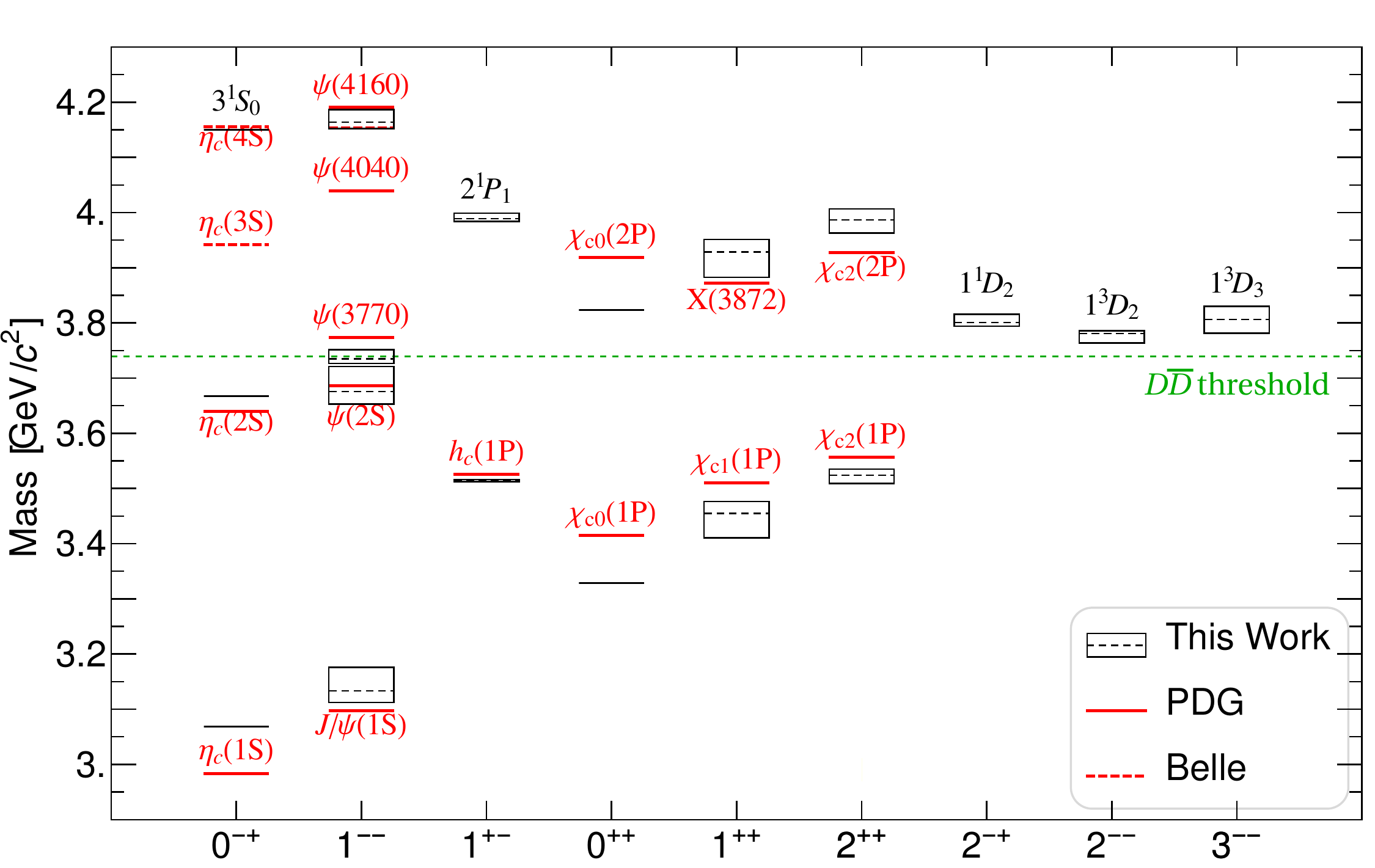}
\includegraphics[width=.489\textwidth]{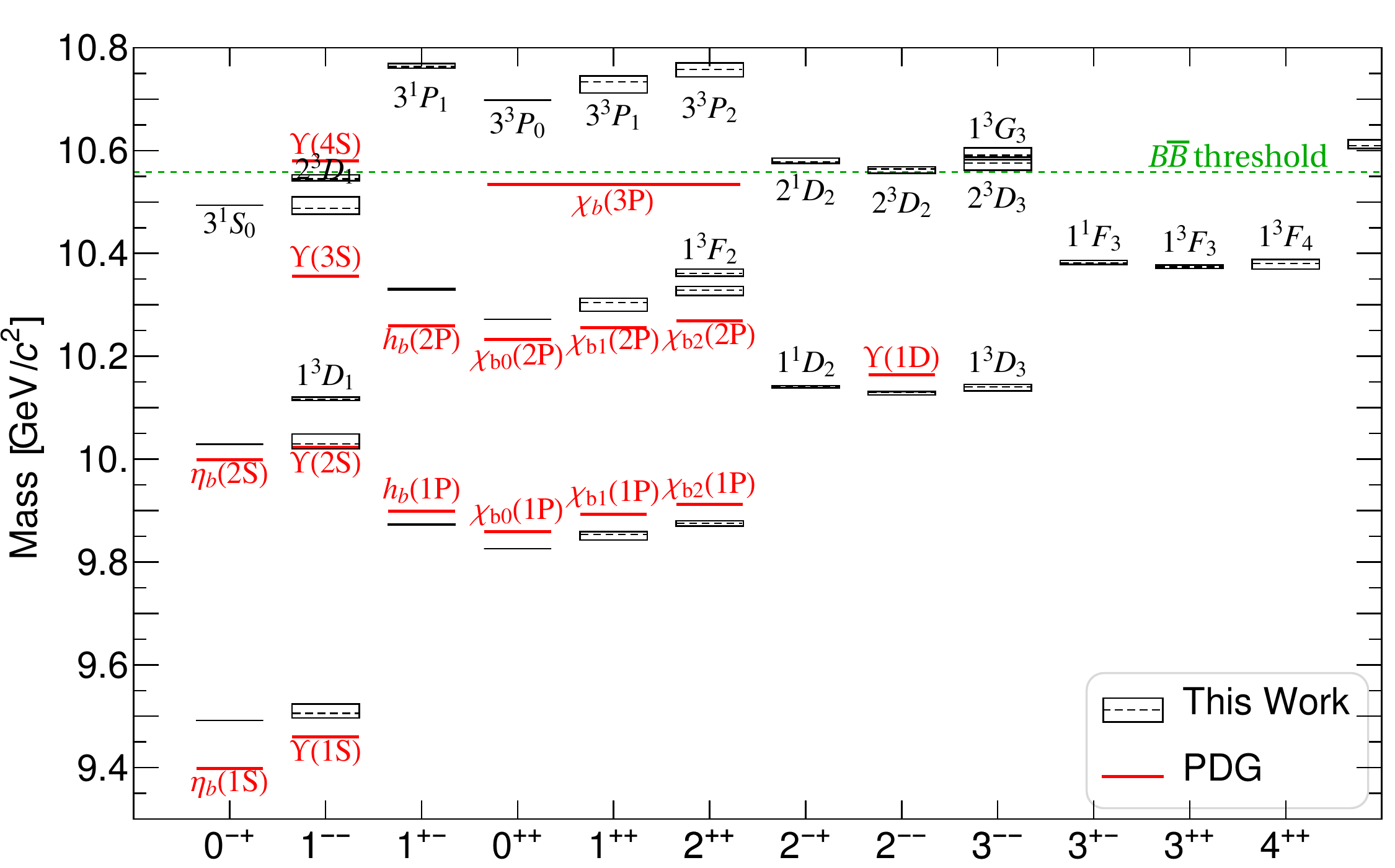}
\caption{\textit{Left}: the charmonium spectrum in GeV;  \textit{Right}: the bottomonium spectrum in GeV. The horizontal axes is
$J^{PC}$. The known states are labeled by their PDG symbols. The unknown states are labeled by the non-relativistic symbols. The spread of
$M_{m_J}$ is indicated by a box and $\overline{M}$ is shown in dashed lines. Model parameters for our spectra shown here are given in
Table~\ref{tab:model_parameters} ($N_{\max}=L_{\max}=24$).
}
\label{fig:quark_spectrum}
\end{figure}

\begin{table}[t]
 \centering
\caption{Summary of the model parameters. The coupling $\alpha_s$ and the gluon mass $\mu_g$ are fixed (see the text). 
The confining strength $\kappa$ and the quark mass $m_q$ are fitted with $m_J=0$ using the experimental data below the 
$D \overline{D}$ or $B \overline{B}$ threshold. The r.m.s. deviations for the $m_J=0$ spectrum, $\delta M_{m_J=0}$, and 
the r.m.s. average-$m_J$ spectrum, $\delta \overline{M}$, are computed for states below the threshold. }
\begin{tabular}{ccc ccc cc} 
\toprule 
   & $\alpha_s$ & $\mu_\text{g}$ (GeV) & $\kappa$ (GeV) & $m_q$ (GeV) & $\delta M_{m_J=0}$ (MeV) & $\delta \overline{M}$ (MeV)  &
$N_{\max}=L_{\max}$ \\ 
\midrule
      $c\bar c$  &  0.3595  & \multirow{2}{*}{0.02} & 0.950 & 1.510 & 64 (8 states) & 52 (8 states) &
\multirow{2}{*}{16} \\
      $b\bar b$ &   0.25  &                       & 1.491 & 4.761 & 56 (14 states)& 51 (14 states) & \\
\midrule 
      $c\bar c$  &  0.3595  & \multirow{2}{*}{0.02} & 0.938 & 1.522 & 65 (8 states) & 52 (8 states) & \multirow{2}{*}{24} \\
      $b\bar b$  &  0.25  &                       & 1.490 & 4.763 & 54 (14 states)& 50 (14 states)&  \\

\bottomrule 
\end{tabular}
\label{tab:model_parameters}
\end{table}
Figure~\ref{fig:quark_spectrum} shows the charmonium and bottomonium spectrum for $N_{\max}=L_{\max}=24$.
As mentioned, the 
mass degeneracy for
$m_J$ is lifted due to the violation of the rotational symmetry by the Fock sector truncation. We use a box to indicate
the spread of masses $M_{m_J}$ from different $m_J$. The r.m.s. values $\overline M=[{(M^2_{-J}+M^2_{1-J}+\cdots 
+M^2_{+J})/(2J+1)}]^\frac{1}{2}$ are shown as dashed bars \cite{Mangin-Brinet03.055203}. We compare our results with the experimental data 
from the 
particle data group (PDG, \cite{pdg.2014}, cf.~\cite{Belle.2014}). The r.m.s. deviations are computed for charmonium (bottomonium)
below the $D\overline{D}$ ($B\overline{B}$) threshold (see Table.~\ref{tab:model_parameters}).
Comparing with the recent results in Ref.~\cite{Gutsche14.096007}, our approach improves the charmonium and bottomonium mass spectra
from light-front holography.

The LFWFs can be used to calculate the transition amplitudes. Here we consider the decay constants. These quantities are
useful
for computing the decay widths and constraining the Standard Model parameters. In the non-relativistic limit, they are proportional to the
wavefunctions at the origin and, therefore, test the short-distance physics of the model. The decay constants for a scalar $S$,
a pseudo-scalar $P$, a axial-vector $A$, and a vector $V$ states are defined as, 
\begin{linenomath*}
\begin{equation}
\begin{split}
 & \langle 0 | \overline{\psi} \gamma^\mu \gamma^5 \psi|P(p)\rangle = ip^\mu f_P ,& \quad 
\langle 0 | \overline{\psi} \gamma^\mu \psi | V(p, \lambda) \rangle = e^\mu_\lambda(p) m_V f_V \, , \quad (\lambda=0,\pm1) \\
 & \langle 0 | \overline{\psi} \gamma^\mu \psi |S(p)\rangle = p^\mu f_S ,& \quad 
\langle 0 | \overline{\psi} \gamma^\mu \gamma^5 \psi | A(p,\lambda) \rangle = e^\mu_\lambda(p) m_A f_A \, , \quad (\lambda=0,\pm1) \\
\end{split}
\end{equation}
\end{linenomath*}
respectively, where $e^\mu_\lambda(p)$ is the spin vector for the vector boson, 
\begin{equation}
 e_\lambda(p) = (e^+_\lambda(p),  e^-_\lambda(p),  \bm e^\perp_\lambda(p)) = 
\left\{ 
\begin{array}{ll}
 \big(\frac{p^+}{m}, \frac{\bm p_\perp^2-m^2}{mp^+}, \frac{\bm p^\perp}{m} \big), &\lambda = 0 \\
 \big(0, \frac{2\bm \epsilon^\perp_\lambda \cdot \bm p^\perp}{p^+}, \bm \epsilon^\perp_\lambda \big), &\lambda = \pm 1. \\
\end{array}\right. 
\end{equation}
Here $\bm \epsilon^\perp_\pm = (1,\pm i)/\sqrt{2}$. Note that due to the charge conjugation symmetry, the decay
constants of $0^{++}$ and $1^{+-}$ vanish. In LFD, the decay constants can be
computed from the ``+'' component of the current
matrix elements, which read \cite{Lepage80.2157}, 
\begin{linenomath*}
\begin{equation}\label{eqn:lf_p_decay_constant}
\begin{split}
f_{P,A} = & 2\sqrt{N_c} \int_0^1\frac{\dd x}{2\sqrt{x(1-x)}} \int\frac{\dd^2 k_\perp}{(2\pi)^3}
\big[\psi^{J}_{m_J=0}(\bm k_\perp, x,\uparrow,\downarrow) - \psi^{J}_{m_J=0}(\bm k_\perp, x,\downarrow,\uparrow) \big], \\
f_{S,V} = & 2\sqrt{N_c} \int_0^1\frac{\dd x}{2\sqrt{x(1-x)}} \int\frac{\dd^2 k_\perp}{(2\pi)^3}
\big[\psi^{J}_{m_J=0}(\bm k_\perp, x,\uparrow,\downarrow) + \psi^{J}_{m_J=0}(\bm k_\perp, x,\downarrow,\uparrow) \big]. 
\end{split}
\end{equation}
\end{linenomath*}
\begin{figure}
 \centering 
\includegraphics[width=0.7\textwidth]{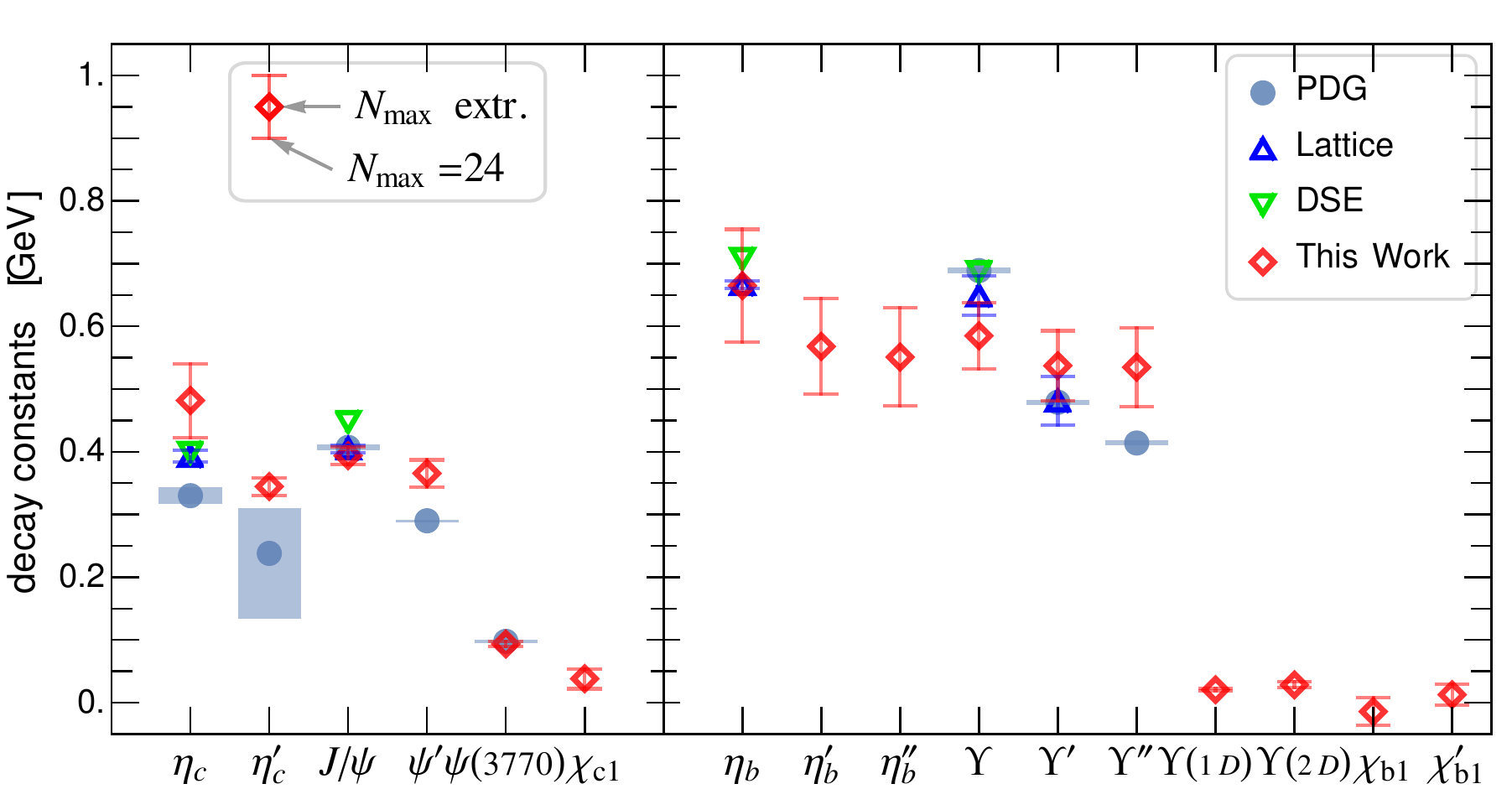}
\caption{The decay constants of scalar and vector quarkonia as compared with PDG data \cite{pdg.2014} as well as Lattice QCD (Lattice)
\cite{Davies10.114504,McNeile12.074503,Donald12.094501,Colquhoun15.074514} and Dyson-Schwinger (DSE) \cite{Blank11.096014}
approaches. 
We extrapolate $N_{\max}$ using second-order polynomials in $N_{\max}^{-1}$ and adopt the difference between 
the extrapolated value ($N_{\max}$ extr.) and the $N_{\max}=24$ value as the uncertainty (not including systematic errors).
Note that $\Upsilon(nD)$ are referring to the vector bottomonia $\Upsilon(n{}^3\!D_1)$.
}\label{fig:decay_constants}
\end{figure}
The decay constants from this work are plotted in Fig.~\ref{fig:decay_constants}. We also list the PDG data \cite{pdg.2014}, as well as
results from
Lattice QCD (Lattice, \cite{Davies10.114504,McNeile12.074503,Donald12.094501,Colquhoun15.074514}), and the Dyson-Schwinger equation (DSE,
\cite{Blank11.096014}). The PDG data are extracted from the dilepton decay widths $\Gamma_{ee}$ (for 
vectors) and diphoton decay widths  $\Gamma_{\gamma\gamma}$ (for pseudo-scalars).
We obtain results for three successive sets of basis regulators, $N_{\max}=L_{\max}=8,16,24$, where the parameters
$\kappa$ and $m_q$ are fitted to the mass spectrum separately, and $\alpha_s, \mu_g$ are kept fixed. 
We then extrapolate $N_{\max}$ using simply polynomials in $N_{\max}^{-1}$ and estimate the uncertainty associated with
$N_{\max}$ from the difference between the extrapolated and the $N_{\max}=24$ results.
While the resultant masses are close as we mentioned, the decay constants show noticeable residual
regulator dependence. This may not be a surprise as the decay constant probes the short-distance physics, whereas the basis is
chosen to emulate confinement. We expect slower convergence for the decay constants than the masses. 

From Fig.~\ref{fig:decay_constants}, our calculated decay constants are in reasonable agreement with the 
known experimental measurements as well as Lattice and DSE results. This is encouraging since the measured decay 
constants are not used in our fits.
However, compared to Lattice and DSE results, most of our results are systematically larger than the PDG data.
This is likely due to the systematic errors of our model and can be improved by more realistic models for the LF
Hamiltonian and/or by including higher Fock spaces. 
Fig.~\ref{fig:decay_constants} also includes decay constants for the $D$-wave states. In the non-relativistic limit, these
quantities should vanish. The small but non-vanishing $D$-wave decay constants in our results indicate the mixing of the $S$-wave
component, as expected from the relativistic treatment.

\begin{figure}
 \centering 
 \includegraphics[width=0.48\textwidth]{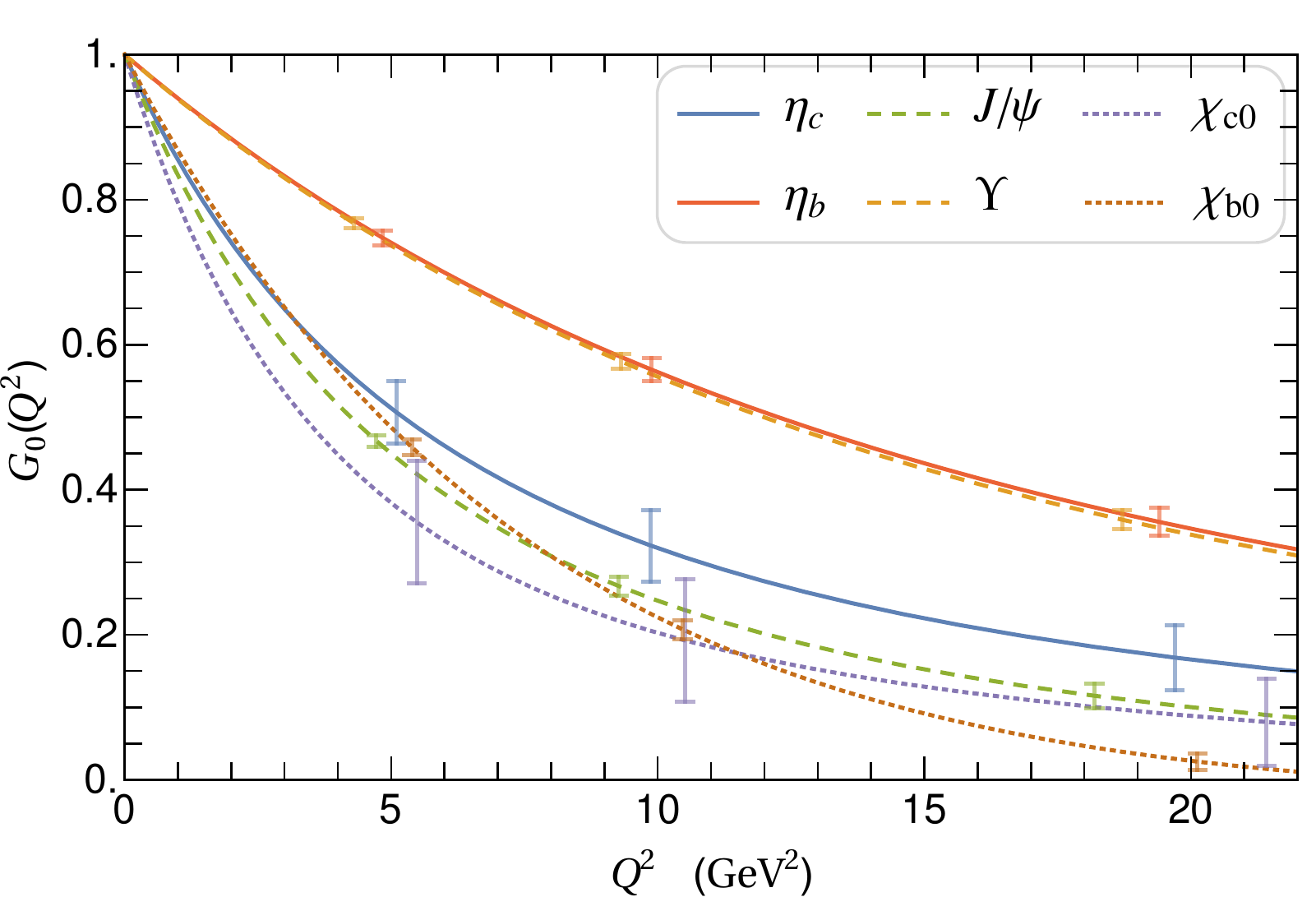}\quad
 \includegraphics[width=0.48\textwidth]{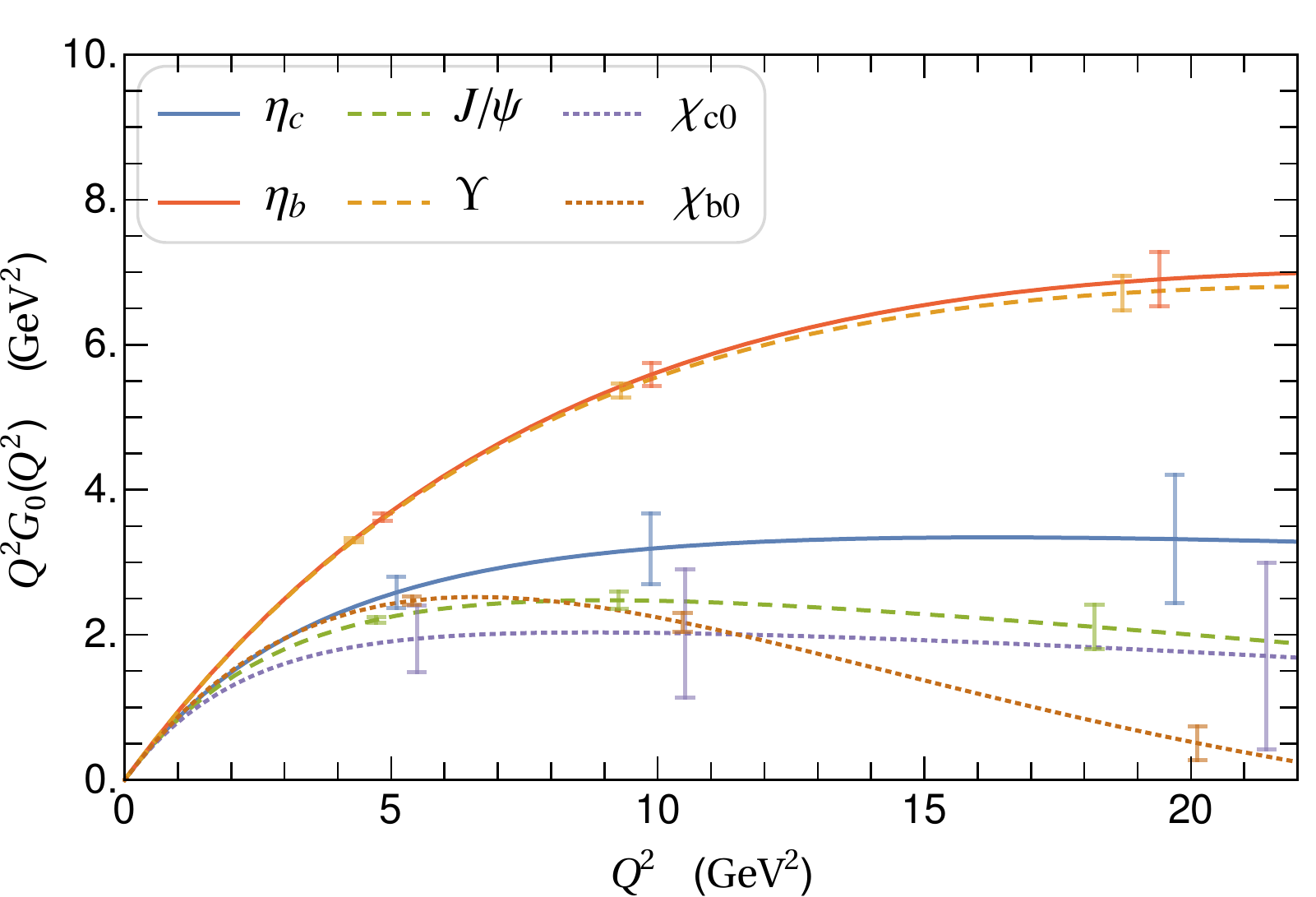}
\caption{The charge form factors $G_0(Q^2)$ for $\eta_c$, $J/\psi$, $\chi_{c0}$, $\eta_b$, $\Upsilon$, and $\chi_{b0}$ 
as extrapolated from $N_{\max}=L_{\max}=8,16,24$ using second-order polynomials in $N_{\max}^{-1}$. The difference between the extrapolated
and the $N_{\max}=24$ values are used to quantify the uncertainty, which does not include systematic errors.
Note that, for the scales we are showing, the charge form factor of $\eta_b$ and $\Upsilon$ are on top of each other.
}
\label{fig:charm_form_factor}
\end{figure}

The LFWFs also provide direct access to other hadronic observables. Here we study the fictitious charge form factor with the photon
coupling only to the quark (but not the anti-quark). Of course, this quantity is not a physical observable.
Nevertheless, it provides important insight to the system. In particular, this charge form factor at small momentum transfer yields the
r.m.s. radius of the hadron,
\begin{linenomath*}
\begin{equation}
  \langle r^2_h \rangle = -6 \frac{\partial }{\partial Q^2} G_0(Q^2)\Big|_{Q\to
0.} %, \quad
%
%magnetic moment
% \mu = \lim_{Q^2\to0} G_1(Q^2), \quad
%
%quadrupole moment
%   \mathsf{Q} = 3\sqrt{2} \frac{\partial }{\partial Q^2} G_2(Q^2)\Big|_{Q\to 0.}
\end{equation}
\end{linenomath*}
In LFD within the impulse approximation, the form factors can be obtained from the Drell-Yan-West formula within the
Drell-Yan frame $P'^+=P^+$ 
\cite{Drell70.181},
\begin{linenomath*}
\begin{equation}
\begin{split}
  I_{m_J,m'_J}(Q^2) 
 \triangleq & \langle \psi^{J^{PC}}_{m'_J}(P')|J^+|\psi^{J^{PC}}_{m_J}(P)\rangle / (2P^+) \\
 =& \sum_{s,\bar s} \int_0^1 \frac{\dd x}{2x(1-x)} \int
\frac{\dd^2k_\perp}{(2\pi)^3} \psi^{J*}_{m'_J}(\bm k_\perp+(1-x)\bm q_\perp, x, s,\bar s) 
\psi^{J}_{m_J}(\bm k_\perp, x,s,\bar s)
\end{split}
\end{equation}
\end{linenomath*}
where $q = P'-P$, and $Q^2 = -q^2 = \bm q^2_\perp$. For (pseudo) scalars, it directly produces the charge form factor
$G_0(Q^2) = I_{0,0}(Q^2)$. For (axial) vector mesons, due to the violation of the rotational symmetry,
there exists some ambiguity on finding the physical (Sachs) form factors $G_0$, $G_1$, and $G_2$ from $I_{m_J,m_J'}$ \cite{Grach84.198}. 
We adopt the prescription of Grach and Kondratyuk \cite{Grach84.198}, which has been shown
to be 
free of zero-mode contributions in some analytical models \cite{Karmanov96.316,deMelo97.2043}. The charge form
factor according to this 
prescription reads,
\begin{linenomath*}
 \begin{equation}
  G_0 = \frac{1}{3}\big[ (3-2\eta) I_{1,1} + 2\sqrt{2\eta}I_{1,0} + I_{1,-1}
\big],
%   G_1 = 2\big[ I_{1,1} - \frac{1}{\sqrt{2\eta}} I_{1,0} \big],
%   G_2 = \frac{2\sqrt{2}}{3}\big[ -\eta I_{1,1} + \sqrt{2\eta} I_{1,0} - I_{1,-1} \big]
 \end{equation}
\end{linenomath*}
where $\eta = Q^2/(4M^2)$, $M$ is the mass of the hadron. Fig.~\ref{fig:charm_form_factor} shows the charge form factors of
$\eta_c$, $J/\psi$, $\chi_{c0}$, $\eta_b$, 
$\Upsilon$, and $\chi_{b0}$.
Table~\ref{tab:rms_radii} lists the r.m.s. radii of the first few states.
From our results, the radius of $J/\psi$ is close to, but slightly larger than, that of $\eta_c$. This is consistent with the Lattice and
DSE results and can be understood from the non-relativistic point of view. Bottomonia are in general smaller than charmonia, 
a result, again, that can be drawn from the non-relativistic argument.
The comparison of the first few available results shows that radii from our approach are in qualitative agreement with 
those of other approaches \cite{Dudek06.074507,Maris07.65}, though ours are systematically smaller than the Lattice and DSE
results. This observation is consistent with the trend of the decay constants. 

\begin{table}
 \centering 
\caption{The mean squared radii (in $\text{fm}^2$) of charmonia and bottomonia as extrapolated from $N_{\max}=L_{\max}=8,16,24$ using
second-order polynomials in $N_{\max}^{-1}$. The difference between the extrapolated and the $N_{\max}=24$ values are presented 
 as the uncertainty, which does not include systematic errors.
}
\label{tab:rms_radii}
\begin{tabular}{lll lll lll}
\toprule
($\text{fm}^2$) & $\langle r^2_{\eta_c} \rangle$ & $\langle r^2_{J/\psi} \rangle$ & $\langle r^2_{\chi_{c0}} \rangle$ & $\langle
r^2_{\eta'_c} \rangle$ & 
$\langle r^2_{\eta_b} \rangle$ & $\langle r^2_{\Upsilon} \rangle$ &
$\langle r^2_{\chi_{b0}} \rangle$ & 
$\langle r^2_{\eta'_b} \rangle$ \\
\midrule 
this work                      & 0.038(5) & 0.0441(8) & 0.06(1) & 0.1488(5) & 0.0146(8) & 0.0149(5) & 0.0331(8) & 0.0510(8) \\
Lattice \cite{Dudek06.074507}   & 0.063(1) & 0.066(2) & 0.095(6) \\
DSE \cite{Maris07.65}           & 0.048(4) & 0.052(3) & \\
\bottomrule
\end{tabular}
\end{table}

\section{Summary and Outlook}\label{sect.5}

We studied heavy quarkonium based on holographic QCD and a realistic one-gluon exchange on the light front. 
We proposed a longitudinal confining potential to incorporate quark mass and longitudinal dynamics. 
We solved the bound-state problem in the Basis Light-Front Quantization approach.
We calculated the spectroscopy, decay constants and charge form factors for the charmonium and bottomonium. 
The comparison with the experimental data and results from Lattice QCD and Dyson-Schwinger equation shows reasonable agreement among the
available observables. As such, we improve the light-front holography from the first approximation to QCD. 

The LFWFs may readily be used in the study of, e.g., radiative transitions \cite{Brambilla11.1534} and diffractive vector meson
production in ultra-peripheral heavy ion collisions \cite{Baltz08.1}. 
These observables will also serve as stringent tests of our model in different regimes. 

This model can also be applied to heavy-light and light-light systems, where the light-front holography
provides a good initial approximation \cite{Brodsky15.1}. 
The phenomenological confinement can be improved by a better understanding of the string/gauge
duality as well as a more complete derivation of the inter-quark potentials from various first-principle approaches to QCD
\cite{Smirnov10.112002}. Ultimately, the phenomenological confining interaction should be replaced by the QCD Hamiltonian.

This work can be extended to higher Fock sectors to incorporate sea-quark and gluon degrees of freedom as well.  The treatment of the 
many-body dynamics is essential for obtaining realistic predictions for states above the thresholds. One of the major challenges of
the light-front Hamiltonian approach however is that explicitly including many gluons, as well as many quark-antiquark pairs, in the Fock
space expansion quickly becomes numerically expensive. The ever increasing computational capacity and the progress in \textit{ab
initio} many-body calculations represent growing opportunities for understanding the strong interaction in the Basis Light-Front
Quantization approach. For systems involving light quarks, however, the na\"ive Fock sector truncation may not suffice, as addressing
dynamical chiral symmetry breaking is essential \cite{Itakura00.045009}. The coherent basis (see, e.g., \cite{Misra00.125017}) and the
light-front coupled-cluster  method \cite{Chabysheva12.417} are two promising approaches for dealing with collective modes and dynamical
symmetry breaking in the light-front Hamiltonian formalism.

\section*{Acknowledgements} 
We wish to thank P. Wiecki, V.A. Karmanov, S.J. Brodsky, G. de T\'eramond, A.P. Trawi\'nski, S. Prell, J.R. Spence,
A.M. Shirokov, S.D. G\l{}azek and G. Chen for valuable discussions. We thank G.~Chen for cross-checking the expressions and the codes.
This work was supported in part by the Department of Energy under
Grant Nos.~DE-FG02-87ER40371 and DESC0008485 (SciDAC-3/NUCLEI) and by
the National Science Foundation under Grant No.~PHY-0904782.
X.~Zhao is supported by the new faculty startup funding by the Institute of Modern Physics, 
Chinese Academy of Sciences under Grant No.~Y532070ZY0.

% \section*{References}



\section*{Supplementary material}

# Supplementary data for "quarkonium in a holographic basis"

Yang Li, Pieter Maris, Xingbo Zhao, and James P. Vary

April 5, 2016

Table 1: Summary of the model parameters. The coupling $\alpha_s$ and the gluon mass $\mu_g$ are fixed. We choose the bottomonium $\alpha_s(9.5\,\text{GeV}) = 0.25$ and then obtain the charmonium coupling $\alpha_s(3.5\,\text{GeV}) \simeq 0.36$ from pQCD evolution of the strong coupling. The confining strength $\kappa$ and the quark mass $m_q$ ($m_q = m_{\bar q}$) are fitted with $m_J = 0$ using the experimental data below the $D\bar{D}$ or $B\bar{B}$ threshold. The r.m.s. deviations for the $m_J=0$ spectrum, $\delta M_{m_J=0}$, and the r.m.s. average-$m_J$ spectrum, $\delta \overline{M}$, are computed for states below the threshold. The numbers for $\kappa$, $m_q$, $\delta M_{m_J=0}$ and $\delta \overline{M}$ are rounded.

| No. | | $\alpha_s$ | $\mu_g$ (GeV) | $\kappa$ (GeV) | $m_q$ (GeV) | $\delta M_{m_J=0}$ (MeV) | $\delta \overline{M}$ (MeV) | $N_{\max} = L_{\max}$ |
|---|---|---|---|---|---|---|---|---|
| I | $c\bar c$ | 0.3595 | 0.02 | 0.963 | 1.492 | 65 (8 states) | 56 (8 states) | 8 |
|   | $b\bar b$ | 0.2500 |      | 1.492 | 4.758 | 59 (14 states) | 55 (14 states) |   |
| II | $c\bar c$ | 0.3595 | 0.02 | 0.950 | 1.510 | 64 (8 states) | 52 (8 states) | 16 |
|    | $b\bar b$ | 0.2500 |      | 1.491 | 4.761 | 56 (14 states) | 51 (14 states) |    |
| III | $c\bar c$ | 0.3595 | 0.02 | 0.938 | 1.522 | 65 (8 states) | 52 (8 states) | 24 |
|     | $b\bar b$ | 0.2500 |      | 1.490 | 4.763 | 54 (14 states) | 50 (14 states) |    |

Table 2: The charmonium spectrum and decay constants ($N_{\max} = L_{\max} = 8, \kappa = 0.963\,\text{GeV}, m_c = 1.492\,\text{GeV}$). $M_{\text{pdg}}$ quotes charmonium masses from Particle Data Group[1] (PDG 2014). $M_{m_J=0}$ collects the masses from the $m_J = 0$ sector. The r.m.s. average-$m_J$ mass $\overline{M} = \left[(M_{-J}^2 + M_{1-J}^2 + \cdots + M_{+J}^2)/(2J+1)\right]^{\frac{1}{2}}$. $M_{\min} = \min\{M_{m_J}\}$, $M_{\max} = \max\{M_{m_J}\}$.

| | $n\,^{2S+1}L_J$ | $J^{PC}$ | $M_{\text{pdg}}$(GeV) | $M_{m_J=0}$(GeV) | $\overline{M}$(GeV) | $M_{\min}$(GeV) | $M_{\max}$(GeV) | $f$ (GeV) |
|---|---|---|---|---|---|---|---|---|
| $\eta_c(1S)^*$ | $1\,^1S_0$ | $0^{-+}$ | 2.9836 | 3.07979 | 3.07979 | 3.07979 | 3.07979 | 0.36081 |
| $J/\psi(1S)^*$ | $1\,^3S_1$ | $1^{--}$ | 3.096916 | 3.14431 | 3.10975 | 3.09233 | 3.14431 | 0.35066 |
| $\chi_{c0}(1P)^*$ | $1\,^3P_0$ | $0^{++}$ | 3.41475 | 3.35285 | 3.35285 | 3.35285 | 3.35285 | 0. |
| $\chi_{c1}(1P)^*$ | $1\,^3P_1$ | $1^{++}$ | 3.51066 | 3.40689 | 3.44322 | 3.40689 | 3.46125 | 0.06155 |
| $h_c(1P)^*$ | $1\,^1P_1$ | $1^{+-}$ | 3.52538 | 3.49972 | 3.49546 | 3.49333 | 3.49972 | 0. |
| $\chi_{c2}(1P)^*$ | $1\,^3P_2$ | $2^{++}$ | 3.5562 | 3.51397 | 3.50420 | 3.48788 | 3.51557 | |
| $\eta_c(2S)^*$ | $2\,^1S_0$ | $0^{-+}$ | 3.6394 | 3.69728 | 3.69728 | 3.69728 | 3.69728 | 0.29471 |
| $\psi(2S)^*$ | $2\,^3S_1$ | $1^{--}$ | 3.686109 | 3.72362 | 3.68679 | 3.66824 | 3.72362 | 0.30062 |
| $\psi(3770)$ | $1\,^3D_1$ | $1^{--}$ | 3.77315 | 3.74542 | 3.72814 | 3.71946 | 3.74542 | 0.07773 |
|  | $1\,^3D_2$ | $2^{--}$ |  | 3.75804 | 3.77571 | 3.75804 | 3.78275 | |
|  | $1\,^1D_2$ | $2^{-+}$ |  | 3.81879 | 3.79676 | 3.78678 | 3.81879 | |
|  | $1\,^3D_3$ | $3^{--}$ |  | 3.83039 | 3.80230 | 3.77322 | 3.83039 | |
| $\chi_{c0}(2P)$ | $2\,^3P_0$ | $0^{++}$ | 3.9184 | 3.88304 | 3.88304 | 3.88304 | 3.88304 | 0. |
| $X(3872)$ | $2\,^3P_1$ | $1^{++}$ | 3.87169 | 3.91125 | 3.95026 | 3.91125 | 3.96962 | 0.07521 |
| $X(3900)^\pm$ | $2\,^1P_1$ | $1^{+-}$ | 3.8887 | 4.01287 | 4.00042 | 3.99419 | 4.01287 | 0. |
| $\chi_{c2}(2P)$ | $2\,^3P_2$ | $2^{++}$ | 3.9272 | 4.01768 | 3.99424 | 3.96949 | 4.01768 | |
|  | $3\,^1S_0$ | $0^{-+}$ |  | 4.20176 | 4.20176 | 4.20176 | 4.20176 | 0.25915 |
| $\psi(4040)$ | $3\,^3S_1$ | $1^{--}$ | 4.039 | 4.21367 | 4.17769 | 4.15958 | 4.21367 | 0.26696 |

\* States whose masses are used to fit the parameters $\kappa$ and $m_c$ in this work.

[1] K.A. Olive et al. (Particle Data Group), Chin. Phys. C **38**, 090001 (2014); [http://pdg.lbl.gov].

Table 3: The charmonium spectrum and decay constants ($N_{\max} = L_{\max} = 16, \kappa = 0.950\,\text{GeV}, m_c = 1.510\,\text{GeV}$). $M_{\text{pdg}}$ quotes charmonium masses from Particle Data Group[1] (PDG 2014). $M_{m_J=0}$ collects the masses from the $m_J = 0$ sector. The r.m.s. average-$m_J$ mass $\overline{M} = \left[(M_{-J}^2 + M_{1-J}^2 + \cdots + M_{+J}^2)/(2J+1)\right]^{\frac{1}{2}}$. $M_{\min} = \min\{M_{m_J}\}$, $M_{\max} = \max\{M_{m_J}\}$.

| | $n\,^{2S+1}L_J$ | $J^{PC}$ | $M_{\text{pdg}}$(GeV) | $M_{m_J=0}$(GeV) | $\overline{M}$(GeV) | $M_{\min}$(GeV) | $M_{\max}$(GeV) | $f$ (GeV) |
|---|---|---|---|---|---|---|---|---|
| $\eta_c(1S)^*$ | $1\,^1S_0$ | $0^{-+}$ | 2.9836 | 3.07306 | 3.07306 | 3.07306 | 3.07306 | 0.40055 |
| $J/\psi(1S)^*$ | $1\,^3S_1$ | $1^{--}$ | 3.096916 | 3.16284 | 3.12261 | 3.10230 | 3.16284 | 0.37249 |

| | $n\,^{2S+1}L_J$ | $J^{PC}$ | $M_{\text{pdg}}$(GeV) | $M_{m_J=0}$(GeV) | $\overline{M}$(GeV) | $M_{\text{min}}$(GeV) | $M_{\text{max}}$(GeV) | $f$ (GeV) |
|---|---|---|---|---|---|---|---|---|
| $\chi_{c0}(1P)^*$ | $1\,^3P_0$ | $0^{++}$ | 3.41475 | 3.33761 | 3.33761 | 3.33761 | 3.33761 | 0. |
| $\chi_{c1}(1P)^*$ | $1\,^3P_1$ | $1^{++}$ | 3.51066 | 3.40861 | 3.44986 | 3.40861 | 3.47030 | 0.05899 |
| $h_c(1P)^*$ | $1\,^1P_1$ | $1^{+-}$ | 3.52538 | 3.51013 | 3.50708 | 3.50555 | 3.51013 | 0. |
| $\chi_{c2}(1P)^*$ | $1\,^3P_2$ | $2^{++}$ | 3.5562 | 3.52568 | 3.51702 | 3.50142 | 3.52823 | |
| $\eta_c(2S)^*$ | $2\,^1S_0$ | $0^{-+}$ | 3.6394 | 3.67956 | 3.67956 | 3.67956 | 3.67956 | 0.32243 |
| $\psi(2S)^*$ | $2\,^3S_1$ | $1^{--}$ | 3.686109 | 3.72234 | 3.67897 | 3.65710 | 3.72234 | 0.33323 |
| $\psi(3770)$ | $1\,^3D_1$ | $1^{--}$ | 3.77315 | 3.74999 | 3.73280 | 3.72418 | 3.74999 | 0.08694 |
| | $1\,^3D_2$ | $2^{--}$ | | 3.76292 | 3.78028 | 3.76292 | 3.78602 | |
| | $1\,^1D_2$ | $2^{-+}$ | | 3.81867 | 3.80099 | 3.79281 | 3.81867 | |
| | $1\,^3D_3$ | $3^{--}$ | | 3.83233 | 3.80674 | 3.77982 | 3.83233 | |
| $\chi_{c0}(2P)$ | $2\,^3P_0$ | $0^{++}$ | 3.9184 | 3.84589 | 3.84589 | 3.84589 | 3.84589 | 0. |
| $X(3872)$ | $2\,^3P_1$ | $1^{++}$ | 3.87169 | 3.89320 | 3.93699 | 3.89320 | 3.95870 | 0.09212 |
| $X(3900)^\pm$ | $2\,^1P_1$ | $1^{+-}$ | 3.8887 | 4.00585 | 3.99503 | 3.98960 | 4.00585 | 0. |
| $\chi_{c2}(2P)$ | $2\,^3P_2$ | $2^{++}$ | 3.9272 | 4.01273 | 3.99120 | 3.96669 | 4.01273 | |
| | $3\,^1S_0$ | $0^{-+}$ | | 4.17128 | 4.17128 | 4.17128 | 4.17128 | 0.29910 |
| $\psi(4040)$ | $3\,^3S_1$ | $1^{--}$ | 4.039 | 4.19738 | 4.15054 | 4.12692 | 4.19738 | 0.28728 |

* States whose masses are used to fit the parameters $\kappa$ and $m_c$ in this work.

[1] K.A. Olive et al. (Particle Data Group), Chin. Phys. C **38**, 090001 (2014); [http://pdg.lbl.gov].

Table 4: The charmonium spectrum and decay constants ($N_{\text{max}} = L_{\text{max}} = 24, \kappa = 0.938\,\text{GeV}, m_c = 1.522\,\text{GeV}$). $M_{\text{pdg}}$ quotes charmonium masses from Particle Data Group[1] (PDG 2014). $M_{m_J=0}$ collects the masses from the $m_J = 0$ sector. The r.m.s. average-$m_J$ mass $\overline{M} = \left[(M_{-J}^2 + M_{1-J}^2 + \cdots + M_{+J}^2)/(2J+1)\right]^{\frac{1}{2}}$. $M_{\text{min}} = \min\{M_{m_J}\}$, $M_{\text{max}} = \max\{M_{m_J}\}$.

| | $n\,^{2S+1}L_J$ | $J^{PC}$ | $M_{\text{pdg}}$(GeV) | $M_{m_J=0}$(GeV) | $\overline{M}$(GeV) | $M_{\text{min}}$(GeV) | $M_{\text{max}}$(GeV) | $f$ (GeV) |
|---|---|---|---|---|---|---|---|---|
| $\eta_c(1S)^*$ | $1\,^1S_0$ | $0^{-+}$ | 2.9836 | 3.06877 | 3.06877 | 3.06877 | 3.06877 | 0.42302 |
| $J/\psi(1S)^*$ | $1\,^3S_1$ | $1^{--}$ | 3.096916 | 3.17612 | 3.13355 | 3.11205 | 3.17612 | 0.37965 |
| $\chi_{c0}(1P)^*$ | $1\,^3P_0$ | $0^{++}$ | 3.41475 | 3.32860 | 3.32860 | 3.32860 | 3.32860 | 0. |
| $\chi_{c1}(1P)^*$ | $1\,^3P_1$ | $1^{++}$ | 3.51066 | 3.41041 | 3.45460 | 3.41041 | 3.47649 | 0.05406 |
| $h_c(1P)^*$ | $1\,^1P_1$ | $1^{+-}$ | 3.52538 | 3.51612 | 3.51375 | 3.51256 | 3.51612 | 0. |
| $\chi_{c2}(1P)^*$ | $1\,^3P_2$ | $2^{++}$ | 3.5562 | 3.53182 | 3.52396 | 3.50908 | 3.53486 | |
| $\eta_c(2S)^*$ | $2\,^1S_0$ | $0^{-+}$ | 3.6394 | 3.66753 | 3.66753 | 3.66753 | 3.66753 | 0.33044 |
| $\psi(2S)^*$ | $2\,^3S_1$ | $1^{--}$ | 3.686109 | 3.72126 | 3.67578 | 3.65283 | 3.72126 | 0.34403 |
| $\psi(3770)$ | $1\,^3D_1$ | $1^{--}$ | 3.77315 | 3.75121 | 3.73492 | 3.72675 | 3.75121 | 0.08943 |
| | $1\,^3D_2$ | $2^{--}$ | | 3.76390 | 3.78070 | 3.76390 | 3.78569 | |
| | $1\,^1D_2$ | $2^{-+}$ | | 3.81578 | 3.80081 | 3.79382 | 3.81578 | |
| | $1\,^3D_3$ | $3^{--}$ | | 3.83024 | 3.80650 | 3.78128 | 3.83024 | |
| $\chi_{c0}(2P)$ | $2\,^3P_0$ | $0^{++}$ | 3.9184 | 3.82321 | 3.82321 | 3.82321 | 3.82321 | 0. |
| $X(3872)$ | $2\,^3P_1$ | $1^{++}$ | 3.87169 | 3.88279 | 3.92842 | 3.88279 | 3.95105 | 0.09723 |
| $X(3900)^\pm$ | $2\,^1P_1$ | $1^{+-}$ | 3.8887 | 3.99890 | 3.98910 | 3.98419 | 3.99890 | 0. |
| $\chi_{c2}(2P)$ | $2\,^3P_2$ | $2^{++}$ | 3.9272 | 4.00641 | 3.98650 | 3.96288 | 4.00641 | |
| | $3\,^1S_0$ | $0^{-+}$ | | 4.15002 | 4.15002 | 4.15002 | 4.15002 | 0.30702 |
| $\psi(4040)$ | $3\,^3S_1$ | $1^{--}$ | 4.039 | 4.18663 | 4.16370 | 4.15218 | 4.18663 | 0.28008 |

* States whose masses are used to fit the parameters $\kappa$ and $m_c$ in this work.

[1] K.A. Olive et al. (Particle Data Group), Chin. Phys. C **38**, 090001 (2014); [http://pdg.lbl.gov].

Table 5: The bottomonium spectrum and decay constants ($N_{\text{max}} = L_{\text{max}} = 8, \kappa = 1.492\,\text{GeV}, m_b = 4.758\,\text{GeV}$). $M_{\text{pdg}}$ quotes bottomonium masses from Particle Data Group[1] (PDG 2014). $M_{m_J=0}$ collects the masses from the $m_J = 0$ sector. The r.m.s. average-$m_J$ mass $\overline{M} = \left[(M_{-J}^2 + M_{1-J}^2 + \cdots + M_{+J}^2)/(2J+1)\right]^{\frac{1}{2}}$. $M_{\text{min}} = \min\{M_{m_J}\}$, $M_{\text{max}} = \max\{M_{m_J}\}$.

| | $n\,^{2S+1}L_J$ | $J^{PC}$ | $M_{\text{pdg}}$(GeV) | $M_{m_J=0}$(GeV) | $\overline{M}$(GeV) | $M_{\text{min}}$(GeV) | $M_{\text{max}}$(GeV) | $f$(GeV) |
|---|---|---|---|---|---|---|---|---|
| $\eta_b(1S)^*$ | $1\,^1S_0$ | $0^{-+}$ | 9.398 | 9.50231 | 9.50231 | 9.50231 | 9.50231 | 0.46979 |
| $\Upsilon(1S)^*$ | $1\,^3S_1$ | $1^{--}$ | 9.4603 | 9.52295 | 9.50957 | 9.50288 | 9.52295 | 0.45714 |
| $\chi_{b0}(1P)^*$ | $1\,^3P_0$ | $0^{++}$ | 9.85944 | 9.82548 | 9.82548 | 9.82548 | 9.82548 | 0. |
| $h_b(1P)^*$ | $1\,^1P_1$ | $1^{+-}$ | 9.8993 | 9.86487 | 9.86623 | 9.86487 | 9.86691 | 0. |
| $\chi_{b1}(1P)^*$ | $1\,^3P_1$ | $1^{++}$ | 9.89278 | 9.83976 | 9.84875 | 9.83976 | 9.85325 | 0.02550 |
| $\chi_{b2}(1P)^*$ | $1\,^3P_2$ | $2^{++}$ | 9.91221 | 9.86946 | 9.86803 | 9.86272 | 9.87262 | |
| $\eta_b(2S)^*$ | $2\,^1S_0$ | $0^{-+}$ | 9.999 | 10.04033 | 10.04033 | 10.04033 | 10.04033 | 0.39332 |
| $\Upsilon(2S)^*$ | $2\,^3S_1$ | $1^{--}$ | 10.02326 | 10.05169 | 10.03727 | 10.03005 | 10.05169 | 0.39680 |

| | $n\,^{2S+1}L_J$ | $J^{PC}$ | $M_{\rm pdg}$(GeV) | $M_{m_J=0}$(GeV) | $\overline{M}$(GeV) | $M_{\rm min}$(GeV) | $M_{\rm max}$(GeV) | $f$(GeV) |
|---|---|---|---|---|---|---|---|---|
| | $1\,^3D_1$ | $1^{--}$ | | 10.11386 | 10.11010 | 10.10822 | 10.11386 | 0.01723 |
| $\Upsilon(1D)^*$ | $1\,^3D_2$ | $2^{--}$ | 10.1637 | 10.11785 | 10.12319 | 10.11785 | 10.12463 | |
| | $1\,^1D_2$ | $2^{-+}$ | | 10.13580 | 10.13294 | 10.13177 | 10.13580 | |
| | $1\,^3D_3$ | $3^{--}$ | | 10.13542 | 10.13272 | 10.12568 | 10.13597 | |
| $\chi_{b0}(2P)^*$ | $2\,^3P_0$ | $0^{++}$ | 10.2325 | 10.28144 | 10.28144 | 10.28144 | 10.28144 | 0. |
| $h_b(2P)^*$ | $2\,^1P_1$ | $1^{+-}$ | 10.2598 | 10.32862 | 10.32696 | 10.32614 | 10.32862 | 0. |
| $\chi_{b1}(2P)^*$ | $2\,^3P_1$ | $1^{++}$ | 10.25546 | 10.29200 | 10.30551 | 10.29200 | 10.31226 | 0.03913 |
| $\chi_{b2}(2P)^*$ | $2\,^3P_2$ | $2^{++}$ | 10.26865 | 10.33087 | 10.32467 | 10.31497 | 10.33127 | |
| | $1\,^3F_2$ | $2^{++}$ | | 10.36309 | 10.35457 | 10.34957 | 10.36309 | |
| | $1\,^3F_3$ | $3^{++}$ | | 10.36561 | 10.36731 | 10.36419 | 10.37097 | |
| | $1\,^1F_3$ | $3^{+-}$ | | 10.38030 | 10.37436 | 10.37182 | 10.38030 | |
| | $1\,^3F_4$ | $4^{++}$ | | 10.38223 | 10.37372 | 10.36290 | 10.38223 | |
| | $3\,^1S_0$ | $0^{-+}$ | | 10.51184 | 10.51184 | 10.51184 | 10.51184 | 0.35815 |
| $\Upsilon(3S)^*$ | $3\,^3S_1$ | $1^{--}$ | 10.3552 | 10.51945 | 10.50394 | 10.49617 | 10.51945 | 0.36402 |
| | $2\,^3D_1$ | $1^{--}$ | | 10.54918 | 10.54435 | 10.54193 | 10.54918 | 0.02425 |
| | $2\,^3D_2$ | $2^{--}$ | | 10.55200 | 10.56022 | 10.55200 | 10.56478 | |
| | $2\,^1D_2$ | $2^{-+}$ | | 10.58138 | 10.57365 | 10.57153 | 10.58138 | |
| | $2\,^3D_3$ | $3^{--}$ | | 10.58271 | 10.57122 | 10.55865 | 10.58271 | |
| | $1\,^3G_3$ | $3^{--}$ | | 10.59956 | 10.58520 | 10.57695 | 10.59956 | |
| | $1\,^3G_4$ | $4^{--}$ | | 10.60157 | 10.59853 | 10.58970 | 10.60684 | |
| | $1\,^1G_4$ | $4^{-+}$ | | 10.61546 | 10.60405 | 10.59818 | 10.61546 | |
| | $1\,^3G_5$ | $5^{--}$ | | 10.61732 | 10.60322 | 10.58792 | 10.61732 | |
| $\chi_{b?}(3P)$ | $3\,^3P_0$ | $0^{++}$ | 10.5794 | 10.72027 | 10.72027 | 10.72027 | 10.72027 | 0. |
| | $3\,^3P_1$ | $1^{++}$ | | 10.72703 | 10.74293 | 10.72703 | 10.75087 | 0.04392 |
| | $3\,^1P_1$ | $1^{+-}$ | | 10.77096 | 10.76538 | 10.76259 | 10.77096 | 0. |
| | $3\,^3P_2$ | $2^{++}$ | | 10.77131 | 10.75943 | 10.74662 | 10.77131 | |
| | $4\,^1S_0$ | $0^{-+}$ | | 10.95999 | 10.95999 | 10.95999 | 10.95999 | 0.31175 |
| $\Upsilon(4S)$ | $4\,^3S_1$ | $1^{--}$ | 10.5794 | 10.96411 | 10.95056 | 10.94377 | 10.96411 | 0.29659 |

$^*$ States whose masses are used to fit the parameters $\kappa$ and $m_b$ in this work.

[1] K.A. Olive et al. (Particle Data Group), Chin. Phys. C **38**, 090001 (2014); [http://pdg.lbl.gov].

Table 6: The bottomonium spectrum and decay constants ($N_{\max} = L_{\max} = 16, \kappa = 1.491\,{\rm GeV}, m_b = 4.761\,{\rm GeV}$). $M_{\rm pdg}$ quotes bottomonium masses from Particle Data Group[1] (PDG 2014). $M_{m_J=0}$ collects the masses from the $m_J = 0$ sector. The r.m.s. average-$m_J$ mass $\overline{M} = \left[(M_{-J}^2 + M_{1-J}^2 + \cdots + M_{+J}^2)/(2J+1)\right]^{\frac{1}{2}}$. $M_{\rm min} = \min\{M_{m_J}\}$, $M_{\rm max} = \max\{M_{m_J}\}$.

| | $n\,^{2S+1}L_J$ | $J^{PC}$ | $M_{\rm pdg}$(GeV) | $M_{m_J=0}$(GeV) | $\overline{M}$(GeV) | $M_{\rm min}$(GeV) | $M_{\rm max}$(GeV) | $f$(GeV) |
|---|---|---|---|---|---|---|---|---|
| $\eta_b(1S)^*$ | $1\,^1S_0$ | $0^{-+}$ | 9.398 | 9.49520 | 9.49520 | 9.49520 | 9.49520 | 0.53938 |
| $\Upsilon(1S)^*$ | $1\,^3S_1$ | $1^{--}$ | 9.4603 | 9.52317 | 9.50657 | 9.49826 | 9.52317 | 0.50967 |
| $\chi_{b0}(1P)^*$ | $1\,^3P_0$ | $0^{++}$ | 9.85944 | 9.82577 | 9.82577 | 9.82577 | 9.82577 | 0. |
| $h_b(1P)^*$ | $1\,^1P_1$ | $1^{+-}$ | 9.8993 | 9.86971 | 9.87106 | 9.86971 | 9.87173 | 0. |
| $\chi_{b1}(1P)^*$ | $1\,^3P_1$ | $1^{++}$ | 9.89278 | 9.84213 | 9.85228 | 9.84213 | 9.85735 | 0.01536 |
| $\chi_{b2}(1P)^*$ | $1\,^3P_2$ | $2^{++}$ | 9.91221 | 9.87471 | 9.87324 | 9.86777 | 9.87798 | |
| $\eta_b(2S)^*$ | $2\,^1S_0$ | $0^{-+}$ | 9.999 | 10.03214 | 10.03214 | 10.03214 | 10.03214 | 0.46047 |
| $\Upsilon(2S)^*$ | $2\,^3S_1$ | $1^{--}$ | 10.02326 | 10.04925 | 10.03120 | 10.02216 | 10.04925 | 0.45700 |
| | $1\,^3D_1$ | $1^{--}$ | | 10.11883 | 10.11464 | 10.11254 | 10.11883 | 0.01827 |
| $\Upsilon(1D)^*$ | $1\,^3D_2$ | $2^{--}$ | 10.1637 | 10.12300 | 10.12834 | 10.12300 | 10.12990 | |
| | $1\,^1D_2$ | $2^{-+}$ | | 10.14099 | 10.13816 | 10.13710 | 10.14099 | |
| | $1\,^3D_3$ | $3^{--}$ | | 10.14346 | 10.13838 | 10.13081 | 10.14346 | |
| $\chi_{b0}(2P)^*$ | $2\,^3P_0$ | $0^{++}$ | 10.2325 | 10.27462 | 10.27462 | 10.27462 | 10.27462 | 0. |
| $h_b(2P)^*$ | $2\,^1P_1$ | $1^{+-}$ | 10.2598 | 10.33064 | 10.32905 | 10.32825 | 10.33064 | 0. |
| $\chi_{b1}(2P)^*$ | $2\,^3P_1$ | $1^{++}$ | 10.25546 | 10.28884 | 10.30469 | 10.28884 | 10.31260 | 0.03509 |
| $\chi_{b2}(2P)^*$ | $2\,^3P_2$ | $2^{++}$ | 10.26865 | 10.33363 | 10.32740 | 10.31715 | 10.33452 | |
| | $1\,^3F_2$ | $2^{++}$ | | 10.36803 | 10.35942 | 10.35435 | 10.36803 | |
| | $1\,^3F_3$ | $3^{++}$ | | 10.37062 | 10.37225 | 10.36921 | 10.37571 | |
| | $1\,^1F_3$ | $3^{+-}$ | | 10.38509 | 10.37936 | 10.37683 | 10.38509 | |
| | $1\,^3F_4$ | $4^{++}$ | | 10.38710 | 10.37873 | 10.36791 | 10.38710 | |
| | $3\,^1S_0$ | $0^{-+}$ | | 10.49843 | 10.49843 | 10.49843 | 10.49843 | 0.43906 |
| $\Upsilon(3S)^*$ | $3\,^3S_1$ | $1^{--}$ | 10.3552 | 10.51176 | 10.49099 | 10.48059 | 10.51176 | 0.44228 |
| | $2\,^3D_1$ | $1^{--}$ | | 10.55174 | 10.54460 | 10.54102 | 10.55174 | 0.02241 |
| | $2\,^3D_2$ | $2^{--}$ | | 10.55509 | 10.56334 | 10.55509 | 10.56747 | |
| | $2\,^1D_2$ | $2^{-+}$ | | 10.58458 | 10.57689 | 10.57415 | 10.58458 | |
| | $2\,^3D_3$ | $3^{--}$ | | 10.58627 | 10.57435 | 10.56074 | 10.58627 | |

|  | $n\,^{2S+1}L_J$ | $J^{PC}$ |  |  |  |  |  |  |
|---|---|---|---|---|---|---|---|---|
|  | $1\,^3G_3$ | $3^{--}$ |  | 10.60407 | 10.58968 | 10.58107 | 10.60407 |  |
|  | $1\,^3G_4$ | $4^{--}$ |  | 10.60613 | 10.58410 | 10.54711 | 10.61116 |  |
|  | $1\,^1G_4$ | $4^{-+}$ |  | 10.61971 | 10.60863 | 10.60269 | 10.61971 |  |
|  | $1\,^3G_5$ | $5^{--}$ |  | 10.62171 | 10.60776 | 10.59234 | 10.62171 |  |
| $\chi_{b?}(3P)$ | $3\,^3P_0$ | $0^{++}$ | 10.5794 | 10.70393 | 10.70393 | 10.70393 | 10.70393 | 0. |
|  | $3\,^3P_1$ | $1^{++}$ |  | 10.71532 | 10.73571 | 10.71532 | 10.74589 | 0.04951 |
|  | $3\,^1P_1$ | $1^{+-}$ |  | 10.76921 | 10.76330 | 10.76034 | 10.76921 | 0. |
|  | $3\,^3P_2$ | $2^{++}$ |  | 10.77023 | 10.75765 | 10.74316 | 10.77023 |  |
|  | $4\,^1S_0$ | $0^{-+}$ |  | 10.93561 | 10.93561 | 10.93561 | 10.93561 | 0.41974 |
| $\Upsilon(4S)$ | $4\,^3S_1$ | $1^{--}$ | 10.5794 | 10.94581 | 10.92267 | 10.91109 | 10.94581 | 0.42446 |

* States whose masses are used to fit the parameters $\kappa$ and $m_b$ in this work.

[1] K.A. Olive et al. (Particle Data Group), Chin. Phys. C **38**, 090001 (2014); [http://pdg.lbl.gov].

Table 7: The bottomonium spectrum and decay constants ($N_{\max} = L_{\max} = 24, \kappa = 1.490\,\text{GeV}, m_b = 4.763\,\text{GeV}$). $M_{\text{pdg}}$ quotes bottomonium masses from Particle Data Group[1] (PDG 2014). $M_{m_J=0}$ collects the masses from the $m_J = 0$ sector. The r.m.s. average-$m_J$ mass $\overline{M} = [(M_{-J}^2 + M_{1-J}^2 + \cdots + M_{+J}^2)/(2J+1)]^{\frac{1}{2}}$. $M_{\min} = \min\{M_{m_J}\}$, $M_{\max} = \max\{M_{m_J}\}$.

|  | $n\,^{2S+1}L_J$ | $J^{PC}$ | $M_{\text{pdg}}$(GeV) | $M_{m_J=0}$(GeV) | $\overline{M}$(GeV) | $M_{\min}$(GeV) | $M_{\max}$(GeV) | $f$(GeV) |
|---|---|---|---|---|---|---|---|---|
| $\eta_b(1S)^*$ | $1\,^1S_0$ | $0^{-+}$ | 9.398 | 9.49182 | 9.49182 | 9.49182 | 9.49182 | 0.57499 |
| $\Upsilon(1S)^*$ | $1\,^3S_1$ | $1^{--}$ | 9.4603 | 9.52401 | 9.50588 | 9.49680 | 9.52401 | 0.53223 |
| $\chi_{b0}(1P)^*$ | $1\,^3P_0$ | $0^{++}$ | 9.85944 | 9.82570 | 9.82570 | 9.82570 | 9.82570 | 0. |
| $\chi_{b1}(1P)^*$ | $1\,^3P_1$ | $1^{++}$ | 9.89278 | 9.84297 | 9.85371 | 9.84297 | 9.85907 | 0.00766 |
| $h_b(1P)^*$ | $1\,^1P_1$ | $1^{+-}$ | 9.8993 | 9.87168 | 9.87301 | 9.87168 | 9.87367 | 0. |
| $\chi_{b2}(1P)^*$ | $1\,^3P_2$ | $2^{++}$ | 9.91221 | 9.87680 | 9.87533 | 9.86982 | 9.88009 |  |
| $\eta_b(2S)^*$ | $2\,^1S_0$ | $0^{-+}$ | 9.999 | 10.02867 | 10.02867 | 10.02867 | 10.02867 | 0.49181 |
| $\Upsilon(2S)^*$ | $2\,^3S_1$ | $1^{--}$ | 10.02326 | 10.04909 | 10.02950 | 10.01969 | 10.04909 | 0.48152 |
|  | $1\,^3D_1$ | $1^{--}$ |  | 10.12060 | 10.11636 | 10.11424 | 10.12060 | 0.01891 |
| $\Upsilon(1D)^*$ | $1\,^3D_2$ | $2^{--}$ | 10.1637 | 10.12479 | 10.13013 | 10.12479 | 10.13172 |  |
|  | $1\,^1D_2$ | $2^{-+}$ |  | 10.14277 | 10.13996 | 10.13891 | 10.14277 |  |
|  | $1\,^3D_3$ | $3^{--}$ |  | 10.14525 | 10.14019 | 10.13264 | 10.14525 |  |
| $\chi_{b0}(2P)^*$ | $2\,^3P_0$ | $0^{++}$ | 10.2325 | 10.27171 | 10.27171 | 10.27171 | 10.27171 | 0. |
| $\chi_{b1}(2P)^*$ | $2\,^3P_1$ | $1^{++}$ | 10.25546 | 10.28761 | 10.30453 | 10.28761 | 10.31298 | 0.02967 |
| $h_b(2P)$ | $2\,^1P_1$ | $1^{+-}$ | 10.2598 | 10.33154 | 10.32993 | 10.32912 | 10.33154 | 0. |
| $\chi_{b2}(2P)^*$ | $2\,^3P_2$ | $2^{++}$ | 10.26865 | 10.33473 | 10.32853 | 10.31821 | 10.33575 |  |
|  | $1\,^3F_2$ | $2^{++}$ |  | 10.36949 | 10.36092 | 10.35586 | 10.36949 |  |
|  | $1\,^3F_3$ | $3^{++}$ |  | 10.37208 | 10.37371 | 10.37070 | 10.37714 |  |
|  | $1\,^1F_3$ | $3^{+-}$ |  | 10.38651 | 10.38082 | 10.37831 | 10.38651 |  |
|  | $1\,^3F_4$ | $4^{++}$ |  | 10.38852 | 10.38020 | 10.36941 | 10.38852 |  |
|  | $3\,^1S_0$ | $0^{-+}$ |  | 10.49364 | 10.49364 | 10.49364 | 10.49364 | 0.47294 |
| $\Upsilon(3S)^*$ | $3\,^3S_1$ | $1^{--}$ | 10.3552 | 10.51028 | 10.48765 | 10.47632 | 10.51028 | 0.47155 |
|  | $2\,^3D_1$ | $1^{--}$ |  | 10.55258 | 10.54524 | 10.54157 | 10.55258 | 0.02352 |
|  | $2\,^3D_2$ | $2^{--}$ |  | 10.55602 | 10.56429 | 10.55602 | 10.56834 |  |
|  | $2\,^1D_2$ | $2^{-+}$ |  | 10.58552 | 10.57790 | 10.57517 | 10.58552 |  |
|  | $2\,^3D_3$ | $3^{--}$ |  | 10.58730 | 10.57542 | 10.56180 | 10.58730 |  |
|  | $1\,^3G_3$ | $3^{--}$ |  | 10.60517 | 10.59085 | 10.58226 | 10.60517 |  |
|  | $1\,^3G_4$ | $4^{--}$ |  | 10.60722 | 10.60423 | 10.59558 | 10.61220 |  |
|  | $1\,^1G_4$ | $4^{-+}$ |  | 10.62070 | 10.60975 | 10.60385 | 10.62070 |  |
|  | $1\,^3G_5$ | $5^{--}$ |  | 10.63009 | 10.60955 | 10.59352 | 10.63009 |  |
| $\chi_{b?}(3P)$ | $3\,^3P_0$ | $0^{++}$ | 10.5794 | 10.69830 | 10.69830 | 10.69830 | 10.69830 | 0. |
|  | $3\,^3P_1$ | $1^{++}$ |  | 10.71193 | 10.73392 | 10.71193 | 10.74490 | 0.04780 |
|  | $3\,^1P_1$ | $1^{+-}$ |  | 10.76906 | 10.76307 | 10.76007 | 10.76906 | 0. |
|  | $3\,^3P_2$ | $2^{++}$ |  | 10.77028 | 10.75775 | 10.74307 | 10.77028 |  |
|  | $4\,^1S_0$ | $0^{-+}$ |  | 10.92944 | 10.92944 | 10.92944 | 10.92944 | 0.45811 |
| $\Upsilon(4S)$ | $4\,^3S_1$ | $1^{--}$ | 10.5794 | 10.94290 | 10.91748 | 10.90475 | 10.94290 | 0.45977 |

* States whose masses are used to fit the parameters $\kappa$ and $m_b$ in this work.

[1] K.A. Olive et al. (Particle Data Group), Chin. Phys. C **38**, 090001 (2014); [http://pdg.lbl.gov].

Table 8: The masses, decay constants for pseudo-scalar and (axial-)vector quarkonia[1]. See Table 1 for model parameters. The masses are taken from the r.m.s. average-$m_J$ values $\overline{M}$. The decay constants are computed using the light-front wavefunction of the $m_J = 0$ sector. The decay constants are also extrapolated (extr.) using simple polynomials in $N_{\max}^{-1}$.

| | | mass $\overline{M}$ (MeV), $N_{\max} = 8, 16, 24$ | | | | decay constant (MeV), $N_{\max} = 8, 16, 24$ | | | | |
|---|---|---|---|---|---|---|---|---|---|---|
| | | PDG[2] | I (8) | II (16) | III (24) | PDG[2ab] | I (8) | II (16) | III (24) | extr. |
| $0^{-+}$ | $\eta_c(1S)$ | 2983.6 | 3080 | 3073 | 3069 | 330±13 | 361 | 401 | 423 | 482 |
| | $\eta_c(2S)$ | 3639.4 | 3697 | 3680 | 3668 | $238^{+72}_{-104}$ | 295 | 322 | 330 | 345 |
| | $\eta_b(1S)$ | 9398.0 | 9502 | 9495 | 9492 | | 470 | 539 | 575 | 665 |
| | $\eta_b(2S)$ | 9999.0 | 10040 | 10032 | 10029 | | 393 | 460 | 492 | 568 |
| | $\eta_b(3S)$ | | 10512 | 10498 | 10494 | | 358 | 439 | 473 | 551 |
| $1^{--}$ | $J/\psi$ | 3096.916 | 3110 | 3123 | 3134 | 407±5 | 351 | 372 | 380 | 394 |
| | $\psi(2S)$ | 3686.109 | 3687 | 3679 | 3676 | 290±2 | 301 | 333 | 344 | 366 |
| | $\psi(3770)$ | 3773.155 | 3728 | 3733 | 3735 | 97.7±3 | 78 | 87 | 89 | 94 |
| | $\Upsilon(1S)$ | 9460.3 | 9510 | 9507 | 9506 | 689±5 | 457 | 510 | 532 | 585 |
| | $\Upsilon(2S)$ | 10023.3 | 10037 | 10031 | 10030 | 479±4 | 397 | 457 | 482 | 537 |
| | $\Upsilon(3S)$ | 10355.2 | 10504 | 10491 | 10488 | 414±4 | 364 | 442 | 472 | 535 |
| | $\Upsilon(1^3D_1)$ | | 10110 | 10115 | 10116 | | 17 | 18 | 19 | 21 |
| | $\Upsilon(2^3D_1)$ | | 10544 | 10545 | 10545 | | 24 | 22 | 24 | 28 |
| $1^{++}$ | $\chi_{c1}(1P)$ | 3510.66 | 3443 | 3450 | 3455 | | 62 | 59 | 54 | 38 |
| | $\chi_{b1}(1P)$ | 9892.78 | 9849 | 9852 | 9854 | | 26 | 15 | 8 | 0* |
| | $\chi_{b1}(2P)$ | 10255.46 | 10306 | 10305 | 10305 | | 39 | 35 | 30 | 13 |
| | $\chi_{b1}(3P)$ | | 10743 | 10736 | 10734 | | 44 | 50 | 48 | 39 |

* The naïve extrapolation yields a negative value ($-14$ MeV), which is corrected to 0, as the quantity should be positive.

[1] The decay constants of $0^{++}$ (scalar meson) and $1^{+-}$ ($h$ axial vector meson) are exactly zero due to charge conjugation symmetry.

[2] K.A. Olive et al. (Particle Data Group), Chin. Phys. C **38**, 090001 (2014); [http://pdg.lbl.gov].

[a] For pseudo scalars ($0^{-+}$), the decay constants are extracted from the diphoton decay width according to[3–4],

$$\Gamma_{P\to\gamma\gamma} = \frac{\pi}{4}\alpha_{\rm em}^2 e_Q^4 m_P^3 |F_{P\gamma}(0)|^2 = 4\pi e_Q^4 \alpha_{\rm em}^2 \frac{f_P^2}{m_P},$$

where $e_Q$ is the charge number of the constituent quark ($e_Q = 2/3$ for charm quark and $e_Q = -1/3$ for bottom quark), $m_P$ is the mass of the pseudo scalar, $\alpha_{\rm em} = \alpha_{\rm em}(Q)$ is the QED running coupling. For charmonium, we take $\alpha_{\rm em}(3.5\,{\rm GeV}) \simeq 1/134$; for bottomonium, we take $\alpha_{\rm em}(9.5\,{\rm GeV}) \simeq 1/132$. The higher order pQCD correction is assumed to be included in the non-perturbative dynamics of the bound states.

[b] For vector mesons ($1^{--}$), the decay constants are extracted from the dilepton decay width according to[3–4],

$$\Gamma_{V\to e^+e^-} = \frac{4\pi}{3}e_Q^2 \alpha_{\rm em}^2 \frac{f_V^2}{m_V}.$$

where $m_V$ is the mass of the vector meson.

[3] S. Godfrey and N. Isgur, Phys. Rev. D **32**, 189 (1985).

[4] C.T.H. Davies, C. McNeile, E. Follana, G.P. Lepage, H. Na, and J. Shigemitsu (HPQCD collaboration), Phys. Rev. D **82**, 114504 (2010).



\end{document}